\documentclass[referee,useAMS,usenatbib]{mn2e}
\usepackage[utf8]{inputenc}
\usepackage{amsmath}
\usepackage{graphicx}
\usepackage{epsfig}
\usepackage{chngcntr}
\usepackage{textcomp}
\def\br{{\cal B}}
\def\brc{{\cal B}_{\rm c}}
\def\brmxc{{\cal B}_{\rm {c}{max}}}
\def\brmnc{{\cal B}_{\rm {c}{min}}}
\def\bcl{{\cal B}_{\rm cl}}

\def\ga{\Gamma}
\def\mdd{\dot{M}}
\def\md{\dot{\cal M}}
\def\mdc{\dot{\cal M}_{\rm c}}
\def\eg{{\it e.g., }}

\def\ie{{\em i.e., }}
\def\vp{v_{\rm p}}
\def\vpc{v_{\rm pc}}
\def\csc{c_{s{\rm c}}}

\def\bp{B_{\rm p}}
\def\bfi{B_{\phi}}
\def\ap{A_{\rm p}}
\def\qbr{Q_{\rm br}}
\def\qcyc{Q_{\rm cycl}}

\def\roa{\rho_{\rm A}}

\def\bpo{B_{\rm p \circ}}
\def\rg{r_{\rm g}}
\def\rd{r_{\rm d}}
\def\rod{\rho_{\rm d}}
\def\rc{r_{\rm c}}
\def\rcl{r_{\rm cl}}
\def\pmin{P_*}
\def\rsh{r_{\rm sh}}
\def\lsim{\lower.5ex\hbox{$\; \buildrel < \over \sim \;$}}
\def\gsim{\lower.5ex\hbox{$\; \buildrel > \over \sim \;$}}
\newcommand{\RN}[1]{
}

\usepackage{graphicx}
\title[Magnetized accretion]
{Study of magnetized accretion flow with variable $\Gamma$ equation of state}
\author[Singh \& Chattopadhyay]
{Kuldeep Singh$^{1,2}$, Indranil Chattopadhyay$^{1}$\thanks{Email:
kuldeep@aries.res.in (KS); indra@aries.res.in (IC)}\\
$^{1}$Aryabhatta Research Institute of Observational Sciences 
(ARIES), Manora Peak, Nainital-263002, India.\\
$^{2}$University of Delhi, Delhi, India.}
\begin{document}
\date{}
\maketitle
\label{firstpage}

\begin{abstract}
We present here the solutions of magnetized accretion flows on to a compact object with hard surface such
as neutron stars. The magnetic field of the central star is assumed dipolar and the magnetic axis is assumed to be aligned with the rotation axis of the star. We have used an equation of state for the accreting fluid
in which the adiabatic index is dependent on temperature and composition of the flow. We have also included cooling processes like bremsstrahlung and cyclotron processes in the accretion flow. We found all possible accretion solutions. All accretion solutions terminate with a shock very near to the star surface and the height of this primary shock do not vary much with either the spin period or the Bernoulli parameter of the flow, although the strength of the shock may vary with the period. For moderately rotating central star there are possible formation of multiple sonic points in the flow and therefore, a second shock far away
from the
star surface may also form. However, the second shock is much weaker than the primary one near the surface.
We found that if rotation
period is below a certain value $(\pmin)$, then multiple critical points or multiple shocks are not possible and
$\pmin$ depends upon the composition of the flow. We also found that cooling effect dominates after the shock and that the cyclotron and the bremsstrahlung cooling processes should be considered
to obtain a consistent accretion solution.   
\end{abstract}

\section{Introduction}
\label{sec:intro}
Magnetic field plays a pivotal role in regulating accretion
on to compact objects like neutron stars (NS) and other magnetized compact objects
with hard surface \citep{pr72,lam73,do73}.
\citet{gl79} showed that the magnetic field can penetrate into the
accretion disc via instabilities and reconnection processes and that there exists a transition
region between the magnetosphere and the accretion disc. \citet{lo86}
presented steady state, axisymmetric magneto-hydrodynamic (MHD) equations of motion in details for a matter flow around a magnetized star with dipolar magnetic field.
\citet{lo95} also studied thin, isothermal disc accretion and possible magnetically driven mass outflow
through open field lines, although the in fall velocity was neglected in the radial
equation of motion.
There are many other studies available in literature for the magnetized
accretion flow on T-Tauri stars and YSOs \citep{cam90,pc96,os95}.

As the accretion disc matter comes closer to the magnetized star, the matter starts to channel
through the magnetic field lines which is known as the funnel flow or accretion curtain.
\citet{li99} studied the funnel flow onto a NS with dipolar field configuration. They assumed
isothermal gas and obtained solution of the Bernoulli integral.
Koldoba et. al. (2002, hereafter KLUR02) followed the equations of motion developed earlier \citep{lo86,u99}
and studied accretion along the `curtain' (i. e., funnel) assuming strong, dipolar  magnetic field configuration and obtained transonic solutions for adiabatic gas.
Karino et. al. (2008, hereafter KKM08) extended \citet{kol02}'s work, but solved for the location of shock in accretion
on to NS. In \citet{ka08}'s study, accretion solutions only match the surface boundary conditions (i.e. flow should have negligible velocity near the star's surface) if shock forms far away from the star.
However,
the terminating shock should be close to the star surface.  In both the studies, the authors used Newtonian gravity and fixed adiabatic index equation of state (EoS) to describe the thermodynamics of the fluid. But, we know that the Newtonian potential fails very close to the compact object and also that the adiabatic index does not remain constant, it varies with temperature
\citep{c38,rcc06}. 

In this paper, we extend the work of \citet{kol02} and \citet{ka08}, by investigating
the funnel flow, using a pseudo-Newtonian potential \citep[hereafter PW potential,][]{pw80} to mimic the strong gravity of NS.
We look for shocks in the accretion flow and we include explicit cooling processes like bremsstrahlung and cyclotron cooling. We use variable adiabatic index $(\Gamma)$ EoS \citep[hereafter CR EoS,][]{cr09} to describe the thermodynamics
of the flow. CR EoS has a composition parameter $(\xi)$ that allows us to study flows with different composition parameter.
We obtain the expression for the Bernoulli integral for a magnetized flow in presence of explicit cooling
and a variable $\Gamma$ EoS.
Similar to \citet{li99,kol02} and \citet{ka08}, we obtain solutions by using the Bernoulli integral as the constant
of motion, but additionally, now this generalized Bernoulli integral is constant even in presence of cooling.
Under these set up, we make a detailed study of all possible accretion solutions through the accretion curtain on to a magnetized compact object. The bulk of the paper describes the solutions while keeping
NS in mind. However, at the end we changed the central object specifications to suite a white dwarf (WD)
and obtain its accretion solutions too.
In section \ref{subsec:MhdEq}, we present the general MHD equations and constants of motion. In
section \ref{subsec:dipmom}, we discuss the strong magnetic field assumption and dipole magnetic field configuration. After this we briefly describe the relativistic EoS having variable adiabatic index in
section \ref{subsec:eos}. The Bernoulli integral which is the total energy, equations of motion and critical point conditions are discussed in section \ref{subsec:berneom}. In the next \S 2.5 we present the reduced MHD shock conditions. The nature of accretion solutions and methodology to solve equations of motion
are explained in section \ref{sec:topo}. In section \ref{sec:result} we present the results. Discussions and concluding remarks are presented in section \ref{sec:conclude}.      

\section{Basic equations and assumptions}
\label{sec:AsumEq}
\subsection{Basic MHD equations}
\label{subsec:MhdEq}
In case of ideal MHD, there are conserved quantities which can be obtained by integrating MHD equations along magnetic field lines by using axis-symmetry assumption. These conserved quantities are labeled by the stream function $\Psi\left(r,\theta \right)$ of the magnetic field.
The MHD equations for steady, inviscid and highly conducting fluids are \citep{h78,lo86,u99},
\begin{equation}
\nabla\ldotp(\rho \textbf{v}) = 0,
\label{cont.eq}
\end{equation}
where, $\rho$ is the mass density and ${\textbf v}\equiv (\vp,0,v_\phi)$ is the velocity vector
and $\vp$ and $v_\phi$ are the poloidal and azimuthal component of the velocity vector.
The equation for no magnetic monopoles,
\begin{equation}
\nabla\ldotp\textbf{B} = 0,
\label{divb.eq}
\end{equation}
${\textbf B}$ is magnetic field. The induction equation,
\begin{equation}
\nabla\times(\textbf{v}\times \textbf{B}) = 0.
\label{induct.eq}
\end{equation}
And the momentum balance equation,
\begin{equation}
(\rho\textbf{v}\ldotp\nabla)\textbf{v} = - \nabla p + \frac{1}{c}(\textbf{J}\times\textbf{B}) + {\bf{\nabla}} \Phi\textbf{\^{r}}.
\label{mblnc.eq}
\end{equation}
In the above equation, ${\textbf J}$ is current density vector and $\Phi$ is the gravitation potential.
In presence of cooling, the $1^{st}$ law of thermodynamics is given by,
\begin{equation}
 \rho \vp \left[\frac{de}{dr}-\frac{p}{\rho^2}\frac{d\rho}{dr} \right]=\qbr+\qcyc=Q,
\label{therm1law.eq}
 \end{equation}
where,  $e=\bar{e}/\rho$ is the internal energy,
${\bar e}$ is the energy
density and $Q$ is the total cooling. $\qbr$ is the bremsstrahlung cooling term
and is given by,
\begin{equation}
\qbr=\lambda_{\rm br}\rho^{2}T_e^{1/2}.
\label{colbr.eq}
\end{equation}
$\qcyc$ is the cyclotron cooling term. Cyclotron cooling is a very complicated process where emission
and resonant absorption both can be important. Therefore, depending on the frequency of the radiation
the flow might behave as an optically thick or thin medium, although the Thompson scattering
optical depth of the flow might be well below one. Generally, such complications are avoided by considering
a cooling function which mimics all the complicated cooling processes \citep{sax98}.
We choose the form of $\qcyc$ given by \citet{cb15},
\begin{equation} 
\qcyc=\lambda_{\rm cycl}\left(\frac{\ap}{10^{15}\mbox{cm}^2}\right)^{-17/40}\left(\frac{\bp}{10^{7}\mbox{G}}\right)^{57/20}\times\left(\frac{\rho}{4\times 10^{-8}\mbox{g}/\mbox{cm}^{3}}\right)^{3/20}\left(\frac{T_e}{10^{8}\mbox{K}}\right).
\label{colcycl.eq} 
\end{equation}
In the above, $\lambda_{\rm br}\sim 5\times 10^{20}\mbox{erg}\mbox{ cm}^{-3}\mbox{g}^{-2}\mbox{s}^{-1}$ and 
$\lambda_{\rm cycl}\sim 1.2\times 10^{8}\mbox{erg}\mbox{ cm}^{-3}\mbox{s}^{-1}$ \citep{cb15} and
$T_e$ is the electron temperature.  

In addition to steady state, we also assume axisymmetry. The conserved quantities are: \\
(i) By integrating continuity equation (\ref{cont.eq}) we obtain the expression of mass flux $(\mdd)$,
\begin{equation}
 \rho \vp\ap = \mbox{constant} = \mdd.
 \label{conMp.eq}
\end{equation}
(ii) Integrating equation (\ref{divb.eq}), we obtain the magnetic flux conservation,
\begin{equation}
 \bp\ap = \mbox{constant},
 \label{conBFp.eq}
\end{equation} 
where, $\bp$ is the poloidal magnetic field component and $\ap$ is the cross-section of the flow orthogonal
to $\bp$.
Using equations (\ref{conMp.eq}) and (\ref{conBFp.eq}), we can write $\vp$ as
\begin{equation}
 \vp=\frac{\kappa(\Psi)}{4\pi\rho}\bp.
 \label{vpBp.eq}
\end{equation} 
(iii) Faraday equation (\ref{induct.eq}) for highly conducting fluid gives the conservation of angular velocity of
field lines $(\Omega)$,
\begin{equation}
 \Omega\left(\Psi\right) = \omega - \frac{\kappa(\Psi) \bfi}{4\pi\rho \varpi} = \mbox{constant},
 \label{conTBp.eq}
\end{equation}
where, $\omega=v_\phi/r$ is the angular velocity and $\varpi=r\rm{sin}\theta$ is the cylindrical radius. \\
(iv) $\phi^{th}$ component of momentum balance equation gives the total angular momentum $(\Lambda)$ which remains conserved along the magnetic field lines,
\begin{equation}
\Lambda(\Psi) = \omega \varpi^{2} - \frac{\bfi\varpi}{\kappa(\Psi)} = \mbox{constant}.
\label{conAngp.eq}
\end{equation}
(v)  Therefore, integration of poloidal component of momentum balance equation gives the
conservation of total energy $(E)$ along the field line,
\begin{equation}
E(\Psi) = \frac{1}{2}{\vp^{2}} + \frac{1}{2}(\omega-\Omega)^{2}\varpi^{2} + h + \Phi(r) - \frac{\Omega^{2}\varpi^{2}}{2}-\int\frac{Qdr}{\rho \vp} = \mbox{constant}.
\label{conEngp.eq}
\end{equation}
In the above equations, $h$ is the specific enthalpy. The form of the PW gravitational potential
is $\Phi(r) = \Phi_{\rm PW}(r) = -{GM}/{(r - \frac{2GM}{c^2})}$
\citep{pw80}. Equation (\ref{conEngp.eq}) is the Bernoulli integral along the magnetic field lines. One can retrieve
the form of Bernoulli integral for the adiabatic flow \citep{lo86,u99,kol02}, if cooling is not considered.
A comparison
of the above, with the Bernoulli integral of black hole accretion disc in the hydrodynamic limit in pseudo-Newtonian potential regime \citep{kscc13, kc14} and in general relativity \citep{ck16} is interesting. In the present case,
angular momentum of the flow ($\omega \varpi^2$), is not conserved due to the effect of
magnetic field, but in the hydrodynamic case it is due to the shear tensor. 

\subsection{Dipole Magnetic field and Assumptions}
\label{subsec:dipmom}
We assume that NS has a dipole-like magnetic field whose magnetic moment $(\mu)$ is aligned with the rotation axis of the star \citep{kol02}. Here, we have assumed that magnetic field is very strong so that the flow does not alter the configuration of the field lines and this also
implies the flow is sub-Alfv\'{e}nic,
\begin{equation}
{\bp^2}/{8\pi}\gg(p,\rho v^{2})~~\mbox{and}~~\roa/\rho\ll 1~~\mbox{or}~~M^{2}_{\rm A}\ll 1.
\label{ass1.eq}
\end{equation}
In the above, the poloidal Alfv\'{e}nic Mach number is defined by
\begin{equation}
M^{2}_{\rm A} \equiv \frac{\vp^{2}}{v^{2}_{\rm Ap}} = \frac{\roa}{\rho},
\label{Mach1.eq}
\end{equation} 
where $v^2_{\rm Ap} = {\bp^2}/{4\pi\rho}$ and $\rho_{\rm A}=4\pi \kappa(\Psi)$. 
The stream function $(\Psi)$ for the dipole magnetic field in spherical coordinates is given by, 
\begin{equation}
\Psi=\frac{\mu}{r}sin^{2}\theta~~\mbox{or}~~r=\rd(\Psi)sin^{2}\theta,
\label{psi1.eq}
\end{equation} where $\rd=\mu/\Psi$ is the radius from where the matter starts channeling the magnetic
field lines and $\bp$ is given by,
\begin{equation}
\bp(r)=\frac{\mu}{r^{3}}(4-3{\rm sin}^2\theta)^{1/2}~~\mbox{or}~~\bp(r)=\frac{\mu}{r^3}(4-3r/\rd)^{1/2}.
\label{bp.eq}
\end{equation}
We can derive the expression for $\omega$ and $\bfi$ by using equation (\ref{conAngp.eq}) and (\ref{conTBp.eq}),
\begin{equation}
\omega = \Omega\frac{1-(\frac{\roa}{\rho})(\frac{\chi}{r^{2}})}{1-\frac{\roa}{\rho}},\bfi = \Omega r\sqrt{4\pi \roa}\frac{1-(\frac{\chi}{r^{2}})}{1-\frac{\roa}{\rho}},
\label{vphip.eq}
\end{equation} where $\chi=\Lambda/\Omega$. If we use assumptions that, $\roa/\rho\ll 1$ and 
$(\roa/\rho)(1-{\chi}/{r^{2}})\ll 1$, then we can obtain relations for $\omega$ and
$\bfi$ (for more detail see \cite{kol02}), 
\begin{equation}
\nonumber
{|\omega - \Omega|}/{\Omega}\ll 1 \mbox{ and } {\bfi}/{\bp}\ll 1.
\end{equation}
The first relation implies that the local angular velocity $(\omega)$ of the fluid remains constant along the field lines and is equal to the angular velocity of the field lines $\Omega(\Psi)$.
Now, if magnetic field lines are frozen into the surface of the star (i.e. no slippage of the
lines) then $\Omega(\Psi)$ can be equated to the angular velocity of the star i.e
$\Omega(\Psi)=\Omega_{\rm star}$. It means that the strong magnetic field forces the matter to
co-rotate with the star.
Therefore, $\rd$ should be close to the co-rotation radius $r_{\rm co}(\equiv \left[{GM_{\circ}}/{\Omega^{2}}\right]^{1/3})$. Our assumption will not work
for $\rd\gg r_{\rm co}$. For that we have to take into account the effect of disc on the
magnetic field configuration. The second relation shows that azimuthal component of magnetic
field $(\bfi)$ is negligible as compared to poloidal component of the magnetic field $(\bp)$,
so we can neglect it.    
 
\subsection{Variable $\Gamma$ EoS}
\label{subsec:eos}
The CR EoS for multispecies flow is given by \citep{cr09}:
\begin{equation}
\bar{e} = \Sigma_{i}e_{i} = \Sigma_{i}\left[n_{i}m_{i}c^{2} 
+ p_{i}\left(\frac{9p_{i} + 3n_{i}m_{i}c^{2}}{3p_{i} + 2n_{i}m_{i}c^{2}}\right)\right],
\end{equation} where, $i$ is the given number of species, $n_i$ is the number density, $m_i$ is the mass and $p_i$ is the pressure. Using the relation for number density, mass density and pressure,
we can simplify the EoS expression and we can obtain the energy density $(\bar{e})$,
\begin{equation}
\bar{e} = n_{e^{-}}m_{e^{-}}c^{2}f(\Theta,\xi) = \rho_{e^{-}}c^{2}f(\Theta,\xi) = \frac{\rho c^{2}f(\Theta,\xi)}{K},
\label{etrnl.eq}
\end{equation} where, $K = [2-\xi(1-1/\eta)]$, $f(\Theta,\xi) = (2-\xi)\left[1 + \Theta\left(\frac{9\Theta + 3}{3\Theta + 2}\right)\right] + \xi\left[\frac{1}{\eta}
+ \Theta\left(\frac{9\Theta + 3/\eta}{3\Theta + 2/\eta}\right)\right]$, $\Theta=\frac{\kappa_{B}T}{m_{e^{-}}c^{2}}$ is dimensionless temperature, $\xi={n_{p}}/{n_{e^{-}}}$, is the composition parameter which is the ratio of number density of proton to the number density of electron.
For $\xi=1.0$, we have only electron-proton plasma. For $0.0<\xi<1.0$ we have electron-positron-proton plasma and for $\xi=0.0$ we have only electron-positron plasma and
$\eta={m_{e^{-}}}/{m_{p}}$ is electron-proton mass ratio. CR is an extremely accurate
approximation of the
relativistic perfect EoS given by \citet{c38}, which contains modified Bessel functions and
hence is difficult to handle in 
numerical calculations. Since CR EoS \ie equation (\ref{etrnl.eq}) is accurate \citep{vkmc15} and yet simple to handle,
we are using this in our analysis.

The enthalpy $h$, variable adiabatic index $\Gamma$ and sound speed $c_{\rm s}$ are given by,
\begin{equation}
h = \frac{\bar{e} + p}{\rho} = \frac{fc^{2}}{K} + \frac{2\Theta c^{2}}{K},
\label{Enthp.eq}
\end{equation} and 
\begin{equation}
\Gamma = 1 + \frac{1}{N},
N = \frac{1}{2}\frac{df}{d\Theta} \mbox{ and } c^{2}_{\rm s}=\frac{2\Theta \Gamma c^{2}}{K}.
\label{gama.eq}
\end{equation}
If we ignore any dissipative processes, then by integrating $1^{st}$ law of thermodynamics without any source/sink
term, we can obtain the adiabatic relation between $\rho$ and $\Theta$ \citep{kscc13},
\begin{equation}{\label{rel_rho.eq}}
\rho={\cal K}\mbox{exp}(k_3) \Theta^{3/2}(3\Theta+2)^{k_1}
(3\Theta+2/\eta)^{k_2},
\end{equation}
where, $k_1=3(2-\xi)/4$, $k_2=3\xi/4$ and $k_3=(f-K)/(2\Theta)$ and ${\cal K}$ is the measure of entropy. 
This relation is equivalent to $p=K\rho^\Gamma$ obtained for a flow described by fixed $\Gamma$ EoS.
Combining equations (\ref{conMp.eq}, \ref{rel_rho.eq}), we obtain the entropy accretion rate $(\dot{\cal M})$
as,
\begin{equation}
\md = \vp\ap\mbox{exp}(k_3) \Theta^{3/2}(3\Theta+2)^{k_1}
(3\Theta+2/\eta)^{k_2}=\mbox{ constant for adiabatic flow}.
{\label{ent_rate.eq}}
\end{equation}
In this paper, we have included radiative cooling processes (equations \ref{colbr.eq}, \ref{colcycl.eq}) unlike \citet{kol02,ka08}, without the consideration of which, we will show later, the flow will not come to rest on the surface of the central compact object.
Since this is not a two temperature solution, so computing the emissivity using $\Theta$ would overestimate
cooling by a large factor. To estimate the electron temperature $T_{e}$, we assume the electron gas
posses the same $N$ as our single temperature, multi-species solution. Therefore, the approximated
electron temperature is given by \citet{kc14}
\begin{equation}
T_{e}=\left[-\frac{2}{3}+\frac{1}{3}\sqrt{4-2\frac{(2N-3)}{(N-3)}}\right]\frac{m_{e^{-}}c^{2}}
{\kappa_{B}}
{\label{electemp.eq}}
\end{equation}

\subsection{Bernoulli function and Equation of motions}
\label{subsec:berneom}
Under the present set of assumptions, the Bernoulli integral from equation (\ref{conEngp.eq}) takes the following form,
\begin{equation}
{\cal B}\left(r,\vp\right)=\frac{\vp^{2}}{2}+h+\Phi_{\rm g}(r)-\int\frac{Qdr}{\rho \vp},
\label{bernl.eq}
\end{equation}
where, $\vp={\mu\kappa(\Psi)}\left(4-3r/\rd\right)^{1/2}/{4\pi\rho r^{3}}$, 
$\Phi_{\rm g}(r)=-{\Omega}^{2}r^{2}_{\rm co}\left(\frac{\alpha \rd}{r-r_{\rm g}} + \frac{r^{3}}{2{\alpha}^{2}\rd^{3}}\right)$, $\alpha={r_{co}}/{\rd}=1$, $\rg={2GM_{\circ}}/{c^{2}}$ and $Q$ is the cooling term.
Gradient of $\vp$ can be obtained by taking the space derivative of the Bernoulli integral (equation \ref{bernl.eq}) ,
\begin{equation}
\frac{d \vp}{dr}=
\frac{{\cal N}(r,\vp,\Theta)}{{\cal D}(r,\vp,\Theta)}.
\label{eom1.eq}
\end{equation}
Where,
\begin{equation}
{\cal N}(r,\vp)=\frac{3c^{2}_{s}}{2r}\left(\frac{8-5r/\rd}{4-3r/\rd}\right)-\frac{\delta}{n}-\Phi^{'}_{\rm g}(r),
\label{numer.eq}
\end{equation}
\begin{equation}
{\cal D}(r,\vp)=\left(\vp-\frac{c^{2}_{s}}{\vp}\right) \mbox{ and } \delta\equiv \frac{Q}{\rho\vp}.
\label{denom.eq}
\end{equation}

We could have considered equation (\ref{ent_rate.eq}) if cooling was not present,
but presently we need to consider the full energy balance (equation \ref{therm1law.eq}),
which gives the gradient in temperature,
\begin{equation}
\frac{d \Theta}{dr}=\left(\frac{\delta K}{2Nc^2}\right)-\frac{\Theta}{N}\left[\frac{1}{\vp}
\frac{d \vp}{dr}+\frac{3\left(8-5r/\rd\right)}{2r\left(4-3r/\rd\right)}\right].
\label{dthdt.eq}
\end{equation}

Flow starts from $\rd$ radius with subsonic velocity i.e $\vp\ll c_{s}$. However, at some critical radius (say $\rc$), the flow becomes transonic  $(\vp=c_{\rm s})$ and is called the sonic point. The sonic point $\rc$ is also the critical point because, at $\rc$ the velocity slope is of the form $d\vp/dr\rightarrow 0/0$. This 
condition gives the critical point relations,
\begin{equation}
{\cal N}(\rc,\vpc,\Theta_{\rm c})={\cal D}(\rc,\vpc,\Theta_{\rm c})=0.
\label{criti.eq}
\end{equation} 
The gradient $(d\vp/dr)_{\rc}$ is obtained with the help of the L'Hospital's rule. So we can
find the slope at the critical point and hence the solution by integration.

\subsection{Shock conditions}
\label{subsec:shock}
The MHD shock conditions \citep{ke89} with the help of strong field approximation reduces to simply hydrodynamics shock conditions, where the information of magnetic field lies inside the poloidal velocity,
\begin{equation}
\nonumber
\left[\rho \vp\right]=0,
\end{equation}
\begin{equation}
\nonumber
\left[\rho \vp^{2} + p\right]=0,
\end{equation}
\begin{equation}
\left[\rho \vp \left\{\frac{\vp^{2}}{2} + h - \int\frac{Qdr}{\rho v_{p}}\right\}\right]=0.
\label{shk.eq}
\end{equation}
Square bracket implies the difference of the pre-shock and post shock flow quantities. The last of the shock
conditions (i.e., conservation of energy flux; equation \ref{shk.eq}) is in principle the flux of
the generalized Bernoulli integral (equation \ref{bernl.eq}).

\section{Methodology}
\label{sec:topo}
In absence of an explicit throat in the flow geometry, it is the presence of gravity (last term in the expression of ${\cal N}$ in eq. \ref{numer.eq}),
which mathematically acts as an effective throat and causes the formation of sonic points or critical points in a flow. Physically, gravity accelerates the inflow from the accretion radius towards the central object
and thereby primarily increases the kinetic energy of the flow. However, since accretion is a convergent flow it increases the temperature as a secondary effect. Therefore, gravity while pulling the flow increases both
the velocity and the temperature (and therefore the sound speed) together, but increases the inflow velocity with a sharper gradient. This causes the flow velocity to cross the local sound speed at the sonic point ($\rc$).
Now if the flow is also rotating, then    
the nature of gravitational interaction can be modified (by centrifugal force), and then the flow can
have multiple critical points (hereafter MCP) \ie the flow
can become transonic at different distances for a given set of parameters. Interestingly,
if the gravity is Newtonian, then the inner sonic point is not formed, but for stronger gravity
like that described by Paczy\'nski-Wiita potential, the inner sonic point is formed. 

Since in this paper, we have considered radiative cooling, so either we have to supply the accretion
rate, or equivalently, supply the density at some distance from the central object. Presently,
we supply the density $\rod$ at $\rd$ or accretion rate $\mdd$. To solve the equations of motion, one need to also specify
the gravitational field via the star's mass $(M_{\circ})$, star's rotation period ($P$) and the surface
magnetic field of the star ($\bpo$). The star radius $R_\circ$ is an input parameter, it is well known
that a small value of $M_{\circ}/R_\circ$ makes $\Phi$
closer to the Newtonian gravitational potential. Our present study is aimed at studying
accretion on to NS,
but we have also modified these input parameters to study funnel accretion onto WD too,
which would be presented later in this paper.
Moreover, rotation period of star and surface magnetic field have some relation as is observed
\citep[see,][]{pan13}. So then, we need only two main parameters, $P$ and $\rod$ (or $\mdd$) as
input parameters and then we can find the poloidal velocity $(v_{\rm pc})$ and dimensionless 
temperature $(\Theta_{\rm c})$ at the critical radius $(\rc)$
from the critical point conditions (equation \ref{criti.eq}).
Therefore, total energy $\brc$ is actually a function of $\rc$.

Flow generally starts with a subsonic velocity from $\rd$ and after passing through a critical/sonic point it becomes supersonic.
The star surface acts as the obstruction for the supersonic matter near the surface and drives a terminating shock. This is true for accretion solutions of all compact objects with hard surface. 
The cooling processes are particularly dominant in the post shock region, since the post-shock is denser and hotter. Apart from the necessity to include cooling from physical arguments, we found that the inclusion of cooling processes is an absolutely necessary condition, in order to obtain a believable accretion solution.
By radiating away a lot of shock heated energy, the cooling processes help to achieve the inner boundary condition of $\vp \rightarrow 0$ as $r \rightarrow R_\circ$. Moreover, the flow is rotating, so multiple sonic points may also occur and if that is so, then the possibility of forming multiple shocks
increases. 

The method to find accretion solution is: 
\begin{enumerate}
\item We find the critical point location and value of flow variables at that point and then velocity slope
$(d\vp/dr)_{\rc}$ with the help of the L'Hospital's rule 
at the critical point for given set of compact object parameters $\bpo,~P,~M_\circ,~R_\circ$ and
supplying the
constant of motion $\brc$.
\item We then integrate the equations forward and backward from the
critical point by using fourth order Runga-Kutta method.
\item While integrating we simultaneously
check for the shock conditions (equation \ref{shk.eq}). If the shock conditions 
are satisfied at some radius $\rsh$ then we compute the post-shock
variables and then start integration from the shock but now on the post shock branch.
\item Near the star surface we search for the post-shock solution which matches the surface boundary condition
of the star.
\end{enumerate}  

\section{Results}
\label{sec:result} 
We have used geometric units, where velocity is in units of $c$ and distance
is in terms of $r_{\rm g}=2GM_{\circ}/c^2$ and mass in units of $M_\circ$. The rotation period ($P$) is quoted
in the text in seconds, although proper conversions were used when feeding into the
equations. In this analysis, for NS we have considered $R_{\circ}=1.0\times 10^{6}$cm and mass $M_{\circ}=1.4M_{\bigodot}$ and the compact object considered in most part of this paper is NS.
Accretion solutions around WD will be explicitly stated.

\begin{figure}
\hspace{3.0cm}
\psfig{figure= 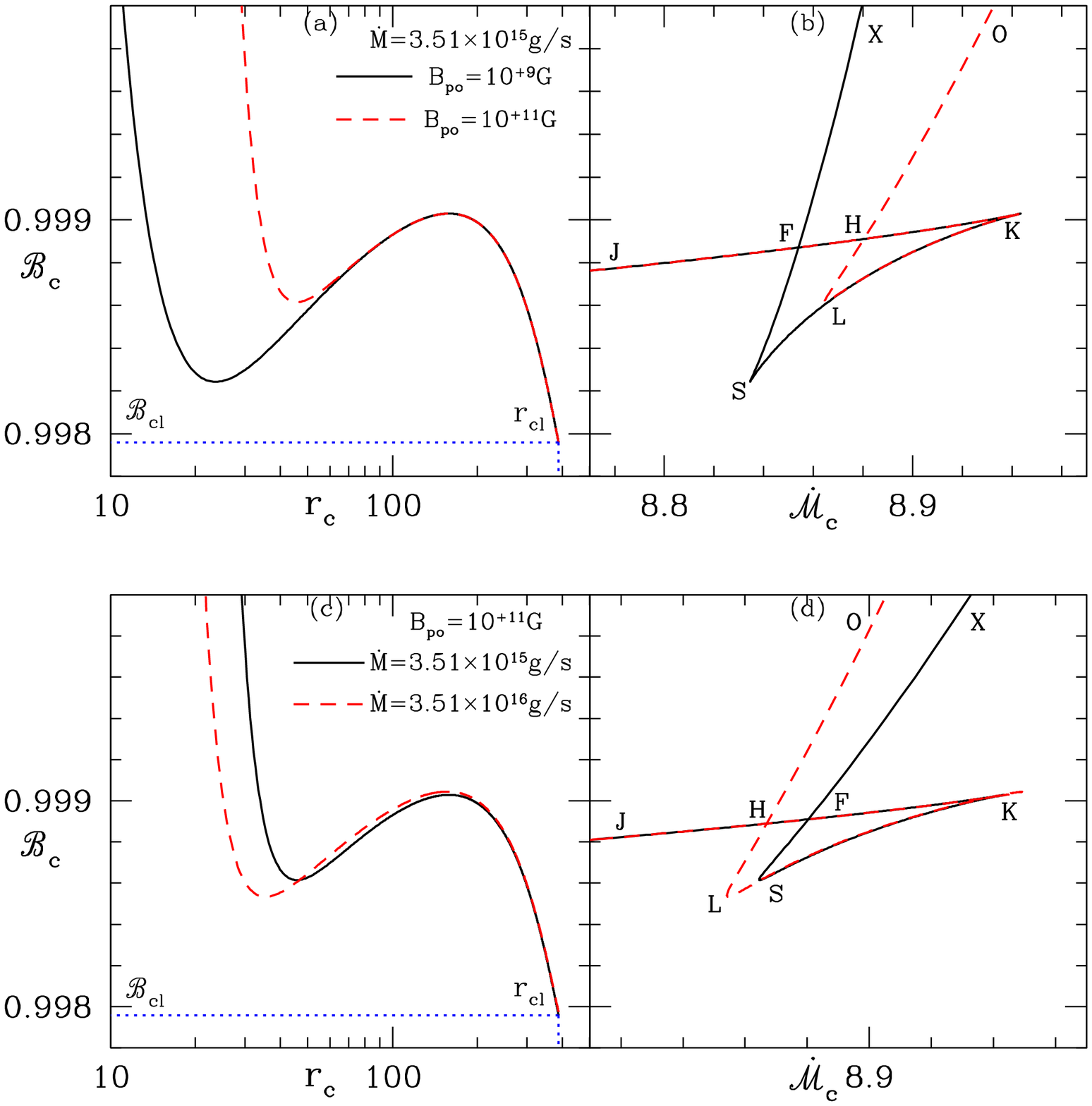,height=9truecm,width=9truecm,angle=0}
\vspace{-0.2cm}
\caption[] {\small Critical energy $\brc$ versus $r_{c}$ (a,c) and $\brc$ versus $\md_{\rm c}$ (b,d). In plots (a, b), solid-black line for $\bpo=10^{9}\mbox{G}$ and dashed-red line for $\bpo=10^{11}\mbox{G}$ with $\mdd = 3.51\times10^{15}\mbox{g}\mbox{ s}^{-1}$.
In plots (c,d), solid-black line for $\mdd = 3.51\times10^{15}\mbox{g}\mbox{ s}^{-1}$
and dashed-red line for $\mdd = 3.51\times10^{16}\mbox{g}\mbox{ s}^{-1}$
with $\bpo=10^{11}\mbox{G}$. All these plots have, P = 1s and $\xi=1$.}
\label{lab:fig1}
\end{figure}

We study the parameter space of MCP using relativistic EoS in the PW potential.
Initially, to construct the parameter space for the critical point conditions
we consider $\bpo$ and $P$ to be independent variables. Other parameter is given by $\mdd$ and in addition we consider an electron-proton flow i.e., $\xi=1.0$. Using equations (\ref{numer.eq}, \ref{denom.eq}) into
equations (\ref{criti.eq}) with the help of definitions in equation (\ref{gama.eq}) we obtain the temperature
at the sonic point ($\Theta_{\rm c}$) as a function of $\rc$ and as a result $\vpc$, $\csc$ and 
$\brc$ can also be expressed as a function of $\rc$. In addition, one can also express the entropy-accretion rate
as a function of $\rc$.
 
We have plotted $\brc$ versus $\rc$ in Figs. (\ref{lab:fig1}a, \ref{lab:fig1}c) and $\brc$ versus $\md_{\rm c}$ in Figs. (\ref{lab:fig1}b, \ref{lab:fig1}d). 
Different curves in the upper two panels (Figs. \ref{lab:fig1}a, \ref{lab:fig1}b) are plotted
for $\bpo=10^{9}\mbox{G}$ (solid, black) and $\bpo=10^{11}\mbox{G}$ (dashed, red), but for the same value of $\mdd = 3.51\times10^{15}\mbox{g}\mbox{ s}^{-1}$.
The curves in the lower two panels
(Figs. \ref{lab:fig1}c, \ref{lab:fig1}d) are plotted for two values of 
$\mdd = 3.51\times10^{15}\mbox{g}\mbox{ s}^{-1}$ 
(solid, black) and $\mdd = 3.51\times10^{16}\mbox{g}\mbox{ s}^{-1}$ 
(dashed-red),
but for the same value of
$\bpo=10^{11}\mbox{G}$. All the plots are for the same $P = 1$s and $\xi=1$.
It may be noted that, each of the $\brc(\rc)$ curve has a maximum $\brmxc$ and a minimum
$\brmnc$. 
Any flow for which $\brmnc\leq\br\leq\brmxc$, there can be three $\rc$, out of which the inner and outer ones are X-type and the middle one is spiral type. 
A flow with $\br\geq \brmxc$ there can be only one inner X-type sonic point. 
Interestingly, the upper limit of $\rc$ is $r_{{\rm cl}}$, for which the
corresponding $\brc$ is  ${ \cal B}_{\rm cl}$. ${ \cal B}_{\rm cl}$ is a function of $P$.
For ${ \cal B}_{\rm cl} \leq \br \leq \brmnc$, only an outer X-type sonic point is possible.
In Figs (\ref{lab:fig1}b, \ref{lab:fig1}d), XS (OL) curve represents the loci of inner critical
points which are X-type, SK (LK) curve represents the loci of middle spiral-type critical points
and KJ curve represents the loci of outer critical points which are also X-type. 
This is the famous `kite-tail' diagram in the energy-entropy space (i. e., $\brc$---$\mdc$ space).
The kite-tail is the enclosed area FSK (or, HLK).
Therefore, if $\brc$ is between the values at coordinate points J and K, then there would three
multiple sonic points. For the range of $\brc$ above that in K, only inner sonic point forms.
If $\brc$ values are in the range between J and S (L), then there are two sonic points, inner
one is X type and outer one is spiral type. These parameters do not produce global solutions.
If $\brc$ is less than the
value at S (L), then for those values of $\brc$ no transonic solution is possible. It must be noted that,
the effect of increasing the surface magnetic field shifts the kite-tail to higher entropy value
(FSK $\rightarrow$ HLK), while by increasing the $\mdd$ 
the kite-tail is shifted to the low entropy region. 
It must also be noted that, if the
gravitational interaction was dictated by Newtonian
potential, then a $\brc$---$\md_{\rm c}$ curve will not have XS (OL) branch, in other words, $\brc$---$\rc$
curves will have a maxima but no minima (see, Fig. \ref{lab:figApp}a, b).
The importance of $\brc$---$\md_{\rm c}$ plot, is to look for the possibility of shock jump between
the inner and outer sonic points. In the range of MCP region, if the inner sonic point is of higher entropy than
the outer sonic point, then there is a possibility of shock transition in accretion, within 
the inner and outer sonic point region. 

\begin{figure}
\hspace{3.0cm}
\psfig{figure= 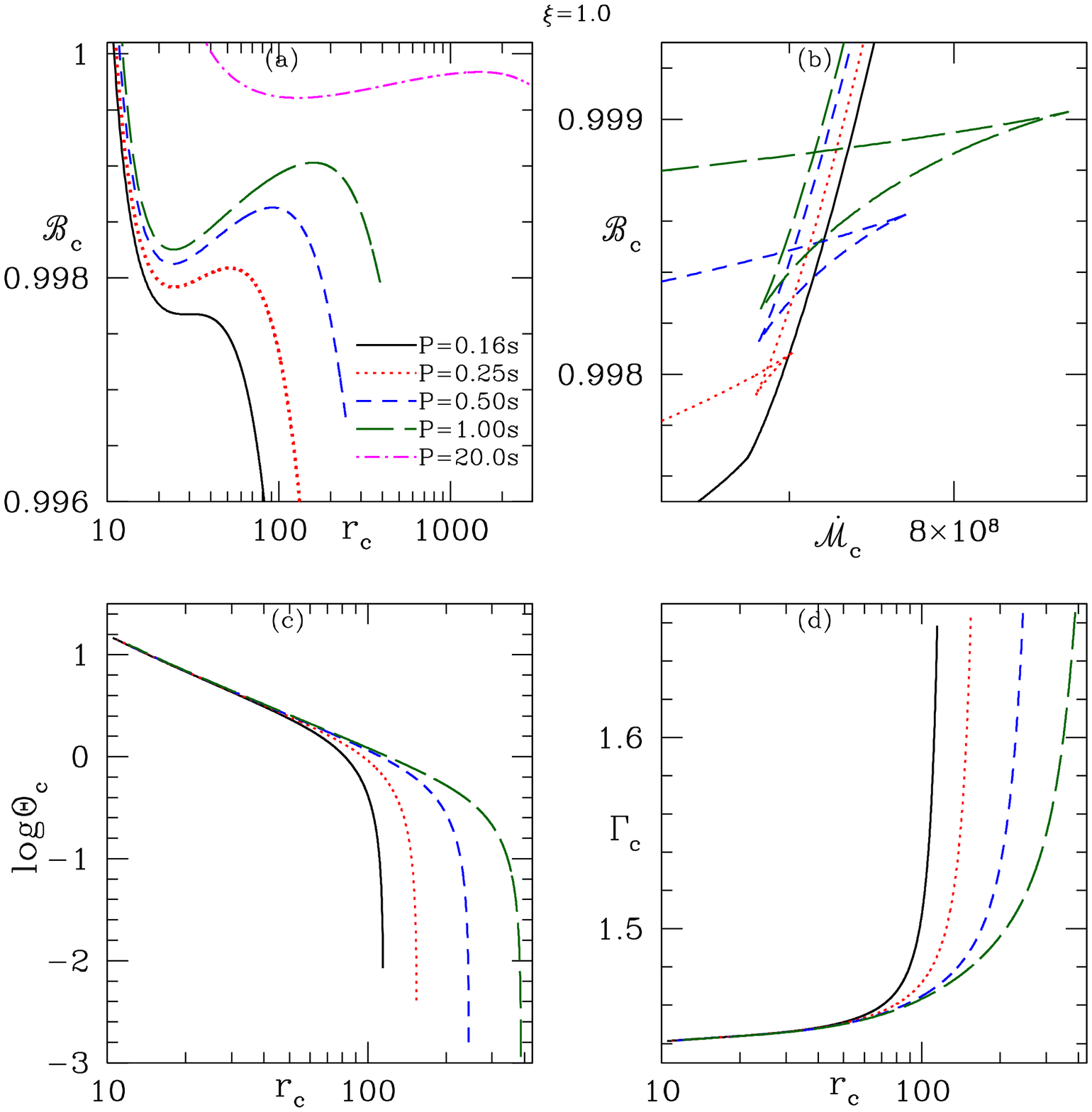,height=9truecm,width=9truecm,angle=0}
\vspace{-0.2cm}
\caption[] {\small (a) $\brc$ as a function of $\rc$, (b) $\brc$ versus $\dot{\cal M}_{c}$, (c) log$\Theta_{\rm c}$ and (d) $(\Gamma_{\rm c})$ versus $\rc$ for rotation periods $P = 0.16$s (Solid, black),
$0.25$s (dotted, red), $0.5$s (dashed, blue), $1.0$s (long-dashed, darkgreen) and $20.0$ (dash-dotted, magenta). For accretion rates, $\mdd_{0.16s} = 0.56\times10^{15}\mbox{g}\mbox{ s}^{-1}$
, $\mdd_{0.25s} = 0.88\times10^{15}\mbox{g}\mbox{ s}^{-1}$,
$\mdd_{0.5s} = 1.8\times10^{15}\mbox{g}\mbox{ s}^{-1}$,
$\mdd_{1.0s} = 3.51\times10^{15}\mbox{g}\mbox{ s}^{-1}$ and 
$\mdd_{20.0s} = 7\times10^{16}\mbox{g}\mbox{ s}^{-1}$. For all the plots 
$\xi=1$.}
\label{lab:fig2}
\end{figure}

As we have commented earlier that, observations of objects containing
NS show a broad correlation between $P$ and $\bpo$ \citep{camilo94,lam05,pan13}. Following,
\citet{pan13}, we assume a simple relation between surface magnetic field and spin
period (in c. g. s units) as $\bpo=10^{0.583log{\rm P}+10}$G, and therefore reduce one
of the input parameters.
By using this relation, we have plotted the $\brc$
(Fig. \ref{lab:fig2}a), $\Theta_{\rm c}$ (Fig. \ref{lab:fig2}c) and $\Gamma_{\rm c}$ (Fig. \ref{lab:fig2}d)
versus $\rc$. Each of the curves were plotted
for rotation periods $P = 0.16$s
(solid, black), $0.25$s (dotted, red), $0.5$s (dashed, blue), $1.0$s (long-dashed, darkgreen)
and $20.0$s (dash-dotted, magenta). All the curves are
plotted for $\mdd_{0.16s} = 0.56\times10^{15}\mbox{g}\mbox{ s}^{-1}$,
$\mdd_{0.25s} = 0.88\times10^{15}\mbox{g}\mbox{ s}^{-1}$,
$\mdd_{0.5s} = 1.8\times10^{15}\mbox{g}\mbox{ s}^{-1}$,
$\mdd_{1.0s} = 3.51\times10^{15}\mbox{g}\mbox{ s}^{-1}$,
$\mdd_{20.0s} = 7\times10^{16}\mbox{g}\mbox{ s}^{-1}$ 
and $\xi=1$.
The $\brc$---$\rc$ plot is quite different from pure hydrodynamic case \citep{kscc13}. In general 
$\brc$ has one $\brmxc$ and one $\brmnc$. Unlike pure hydrodynamic case,
the location of $\brmxc$ and $\brmnc$ approaches each other, with decreasing $P$ (or increasing rotation). Eventually, the maxima and minima merges at $P=0.16$s. The dip between a $\brmxc$ and $\brmnc$ increases
as $P$ increases from $0.16$s --- $1$s. However, for very large value of $P$ ($=20$s), the dip decreases
and finally only monotonic variation of $\brc$ with $\rc$ is possible for very low rotation.
As is expected $\Theta_{\rm c}$ and $\Gamma_{\rm c}$ do not show the presence of extrema.
Since $\Gamma$ is a function of $\Theta$ and $\xi$, $\Gamma_{\rm c}$ is not constant.
In Fig. (\ref{lab:fig2}b), we plot $\brc$ versus $\md_{\rm c}$ were the curves are for the same
values of $P$ as mentioned above. Interestingly, the kite-tail do not form for $P=0.16$s (solid, black),
and starts to form as $P$ is increased. The area encompassed by the kite-tail also increases
as $P\rightarrow 0.25$---$1.0$. However, for very high $P$, the $\brc$---$\mdc$ curve
do not form a kite tail and it opens up.

\begin{figure}
\hspace{3.0cm}
\psfig{figure= 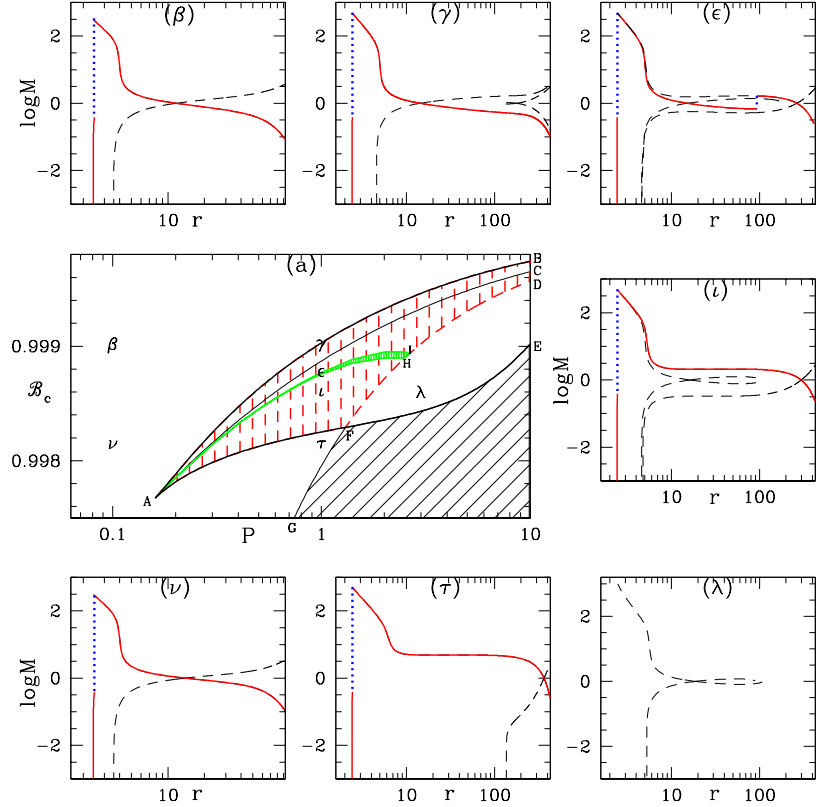,height=11truecm,width=11truecm,angle=0}
\vspace{-0.2cm}
\caption[] {\small (a) Parameter space $\brc- {\rm P}$ shows MCP region bounded by BAFE (solid, black).
The shaded region CAFDC (dashed, red) three $\rc$. Region DFE produces two $\rc$. 
GFD is the curve of $\bcl$ as a function of $P$.
AHIA is the second shock parameter space. The region below GFE (shaded with slanted lines) do not produce transonic solutions. 
Points, $\beta$, $\epsilon$, $\iota$, $\lambda$, $\tau$ and $\nu$ are coordinate points which
represents various regions in $\brc$---$\md_{\rm c}$
space. Corresponding solutions $M$ versus $r$ are plotted in identically named panels ($\beta$---$\nu$).
Accretion solutions (solid, red), shock transitions (dotted, blue), wind type or multi valued solutions (dashed, black). Here, $\mdd_{0.1s} = 0.35\times10^{15}\mbox{g}\mbox{ s}^{-1}$
, $\mdd_{1.0s} = 3.51\times10^{15}\mbox{g}\mbox{ s}^{-1}$ and 
$\mdd_{3.0s} = 1.1\times10^{16}\mbox{g}\mbox{ s}^{-1}$ 
and $\xi=1$.}
\label{lab:fig3}
\end{figure}

One would like to know, for a given ${\dot M}$ what is the range of flow parameters $\br$ and $P$,
for which multiple sonic points are possible and what would be the typical solution for a certain combination
of $\br$ and $P$.
In Fig. (\ref{lab:fig3}a), we plot the locus of $\brmxc$ (AB) and $\brmnc$ (AFE) as a function of $P$
keeping $\rod=10^{-10}$g cm$^{-3}$ same, for an electron-proton flow.
Therefore, flows with any pair of $\br$, $P$ parameters within the bounded region $BAFE$, would
harbour multiple sonic points (two or three). From the $\brc$---$\rc$ plots we have seen that there exists a maximum limit
of sonic points $\rcl$ which corresponds to $\bcl$ in terms of Bernoulli parameter (marked in Figs. \ref{lab:fig1}a, c).
Plotting $\bcl$ as a function of $P$, produces the curve GFD in $\br$---$P$ space (Fig. \ref{lab:fig3}a).
Depending on $P$, $\bcl < \brmnc$, or may also be $\bcl > \brmnc$. Flows with parameters within
the region DFE (\ie where, $\bcl > \brmnc$), have only two sonic
points and do not produce global solutions (i. e., solutions connecting $\rd$ and $R_\circ$). Flow parameters
from the region bounded by GFE (shaded with slanting lines) can never be transonic. Parameters from the region BAFD (shaded with vertical dashed lines), produce flows containing three sonic points and AC (thin black line within BAFD region) is the same entropy line \ie inner and outer critical point
has same entropy.
The thin shaded strip AHIA within the region for three sonic point, harbours second shock at a larger distance
from the star surface. In $\br$---$P$ parameter space, we mark various coordinate points as
$\beta$ ($\br=0.99902$, $P=0.1$s), $\gamma$ ($\br=0.99902,~P=1$s), $\epsilon$ ($\br=0.99877,~P=1$s),
$\iota$ ($\br=0.9986,~P=1$s), $\lambda$ ($\br=0.9986,~P=3$s), $\tau$ ($\br=0.99814,P=1$s) and $\nu$
($\br=0.99814,~P=0.1$s). Here accretion rates are $\mdd_{0.1s} = 0.35\times10^{15}\mbox{g}\mbox{ s}^{-1}$
, $\mdd_{1.0s} = 3.51\times10^{15}\mbox{g}\mbox{ s}^{-1}$ and 
$\mdd_{3.0s} = 1.1\times10^{16}\mbox{g}\mbox{ s}^{-1}$.
Each such point is the representative of the domain in the parameter space.
In Figs. (\ref{lab:fig3}$\beta$---\ref{lab:fig3}$\nu$), starting from top left corner in a clock
wise manner, we plot the accretion solutions \ie Mach number $M=\vp/c_s$ as a function of $r$, corresponding to the coordinate points $\beta$---$\nu$
in Fig. (\ref{lab:fig3}a). The panels are named similar to the coordinate points in $\br$---$P$ parameter space.
The physical accretion solutions (solid, red) connects $\rd$ to $R_\circ$. The dotted (blue) vertical lines represent shock transition. The dashed (black) curve represent either wind type or multi-valued solutions and cannot be considered
proper accretion solutions. In general, the crossing points in the solutions signify the locations
of sonic points $\rc$. For three sonic point region (coordinate points $\epsilon$ and $\iota$), the middle
spiral type sonic point is not shown, but is typically located near the region where $dM/dr \rightarrow 0$ for
upper or lower branch. 
In order to understand how $P$ and $\br$ affect the solutions, we started with the $\beta$ point, kept the same
$\br$ but increased $P$ to reach to the point $\gamma$. Then kept $P$ same and reduced $\br$ to reach $\epsilon$
and $\iota$ and $\tau$.
Then again kept the same $\br$ as $\tau$ but decreased $P$ to reach to $\nu$, where the $P$ of $\nu$ and $\beta$
are same. Point $\beta$ is of higher energy but of low $P$ (\ie high spin). Higher spin causes the matter to rotate faster, and thus significant portion of $\br$ is in the form of rotational energy. Therefore, the flow
could gain enough $\vp$ to become transonic, only when it is closer to the compact object.
Increasing the $P$ (\ie at the point $\gamma$) by a moderate amount, reduces the rotation energy
and therefore the interplay between
gravity and centrifugal terms generate multiple sonic points, although the global solution
is still through the inner sonic point. Reducing the specific energy or $\br$ further, makes the rotational
and gravity terms comparable enough, not only to cause the accretion flow to pass through the outer sonic point,
but can also trigger shock transition between inner and outer sonic points ($\epsilon$).
If the energy is reduced even further, then the flow pressure decreases to the extent such that its combined effect
with rotational energy is lower than that of gravitational pull and hence in spite of the presence of three
sonic points, the second shock do not form.
Point $\lambda$ has the same $\br$ as $\iota$, but has much higher $P$. Such low rotation as well as low energy,
makes the flow non-transonic, \ie transonic solution is not global. Point $\tau$ is outside the
MCP region and of low energy, so there is only one sonic point but far away from the central object.
Keeping the same $\br$ but reducing $P$ cause the sonic point to form closer to the central star.
Since the central object has hard surface, all the global accretion solutions end with a
terminating shock.

\begin{figure}
\hspace{3.0cm}
\psfig{figure= 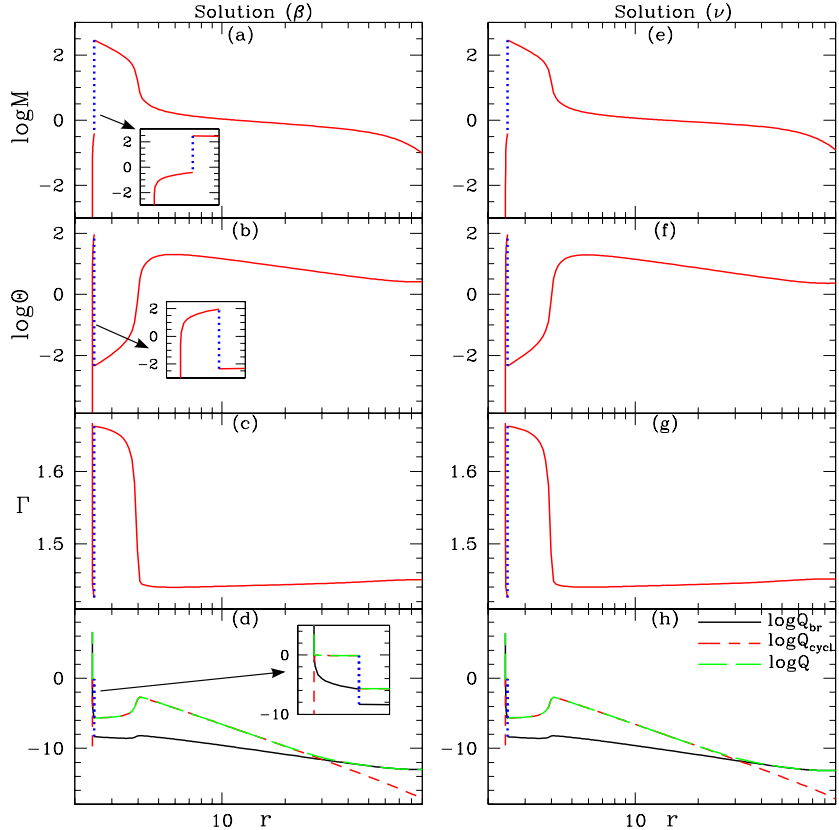,height=11truecm,width=11truecm,angle=0}
\vspace{-0.2cm}
\caption[]{\small Variation of log$M$ (a, e); log$\Theta$ (b, f); $\Gamma$ (c, g);
total radiative losses in log scale $Q$ (long dashed, green), bremsstahlung losses $\qbr$ (solid, black)
and cyclotron losses $\qcyc$ (dashed, red) in panels (d, h) as a function of $r$. Comparison of solutions
corresponding to
coordinate points $\beta$ (a, b, c, d) and $\nu$ (e, f, g, h) of parameter space in Fig. (\ref{lab:fig3}a).
The inset in panels a, b, d zooms the inner shock region. Here, $\mdd_{\beta,\nu} = 0.35\times10^{15}\mbox{g}\mbox{ s}^{-1}$ 
and $\xi=1$.}
\label{lab:fig4}
\end{figure}

To understand
the physics of accretion onto a magnetized compact object, we should compare the distribution of other flow variables in addition to spatial distribution of the $M$.
In the left panels of Fig. (\ref{lab:fig4}), we have plotted the variables log$M$ (Fig. \ref{lab:fig4}a), log$\Theta$ (Fig. \ref{lab:fig4}b), $\Gamma$ (Fig. \ref{lab:fig4}c) and the cooling rates log$Q$, log$\qbr$ and
log$\qcyc$ (Fig. \ref{lab:fig4}d) for the solution corresponding to point $\beta$ in $\br$---$P$ space (Fig. \ref{lab:fig3}a). We compared the same flow variables for the solution corresponding to the point $\nu$ in (Fig. \ref{lab:fig3}a) in the right panels (Fig. \ref{lab:fig4}e-h). The parameters at point $\beta$ and $\nu$ are differentiated
by $\br$ but with the same $P$. The solution type are therefore similar, except that
the sonic point of higher $\br$ solution is located closer to the central star. The terminating shock
is located at similar distance in the two cases, and post-shock flow variables as well as, the cooling rates
are also similar. The radiative efficiency of the pre-shock flow is around $0.05-0.06$ but that of the post-shock flow is about $0.3$.

\begin{figure}
\hspace{3.0cm}
\psfig{figure= 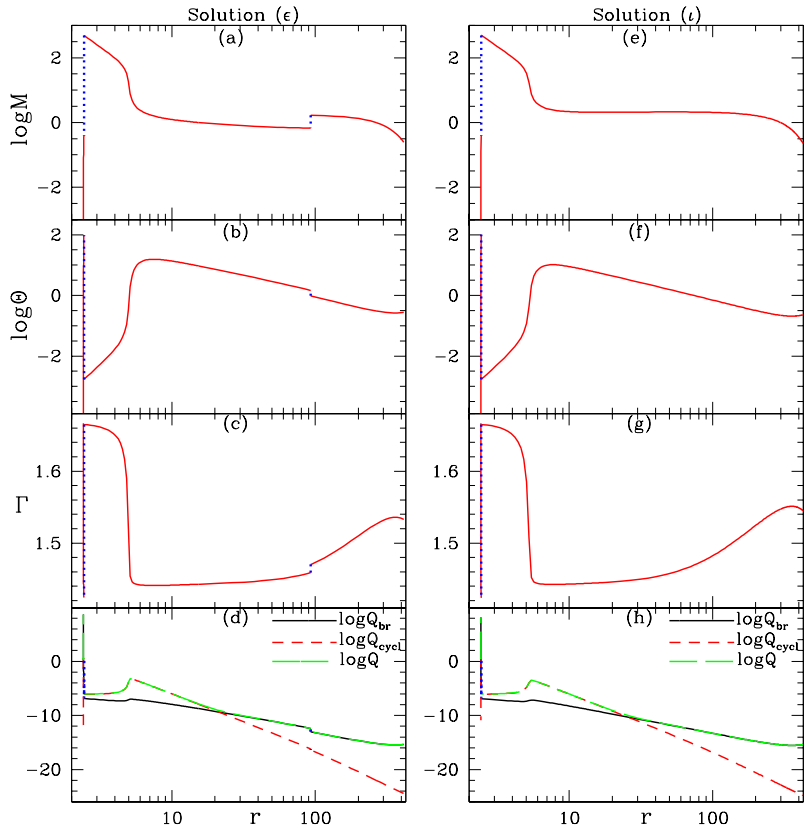,height=11truecm,width=11truecm,angle=0}
\vspace{-0.2cm}
\caption[] {\small Variation of  log$M$ (a, e); log$\Theta$ (b,f); $\Gamma$ (c,g);
total radiative losses in log scale $Q$ (long dashed, green), bremsstrahlung losses $\qbr$ (solid, black)
and cyclotron losses $\qcyc$ (dashed, red) in panels (d, h) as a function of $r$. Comparison of solutions
corresponding to
coordinate points $\epsilon$ (a, b, c, d) and $\iota$ (e, f, g, h) of parameter space in Fig. (\ref{lab:fig3}a).
Here, $\mdd_{\epsilon,\iota} = 3.51\times10^{15}\mbox{g}\mbox{ s}^{-1}$ 
and $\xi=1$.}
\label{lab:fig5}
\end{figure}

In Fig. (\ref{lab:fig5}), we compare flow variables of two solutions in the parameter space range for
three sonic points. On the left panels, we plot log$M$ (Fig. \ref{lab:fig5}a), log$\Theta$ (Fig. \ref{lab:fig5}b), $\Gamma$ (Fig. \ref{lab:fig5}c) and the cooling rates log$Q$, log$\qbr$ and log$\qcyc$ (Fig. \ref{lab:fig5}d) for the solution corresponding to point $\epsilon$ in $\br$---$P$ space (Fig. \ref{lab:fig3}a).
This solution harbours two shocks (dotted, blue vertical line). In the right panels, we plot
the same corresponding variables (Fig. \ref{lab:fig5}e-h), but now for the parameters which 
characterize coordinate point $\iota$ in the parameter space of Fig. (\ref{lab:fig3}a).
This set of parameters also produce three sonic points, but shock condition for the
second shock is not satisfied. The temperature of the two shocked solution (corresponding to $\epsilon$)
is slightly higher, and the $Q$ is also slightly higher than the one shock solution (corresponding to $\iota$).
Second shock is noticeably weaker than the terminating shock close to the star surface. 

\begin{figure}
\hspace{3.0cm}
\psfig{figure= 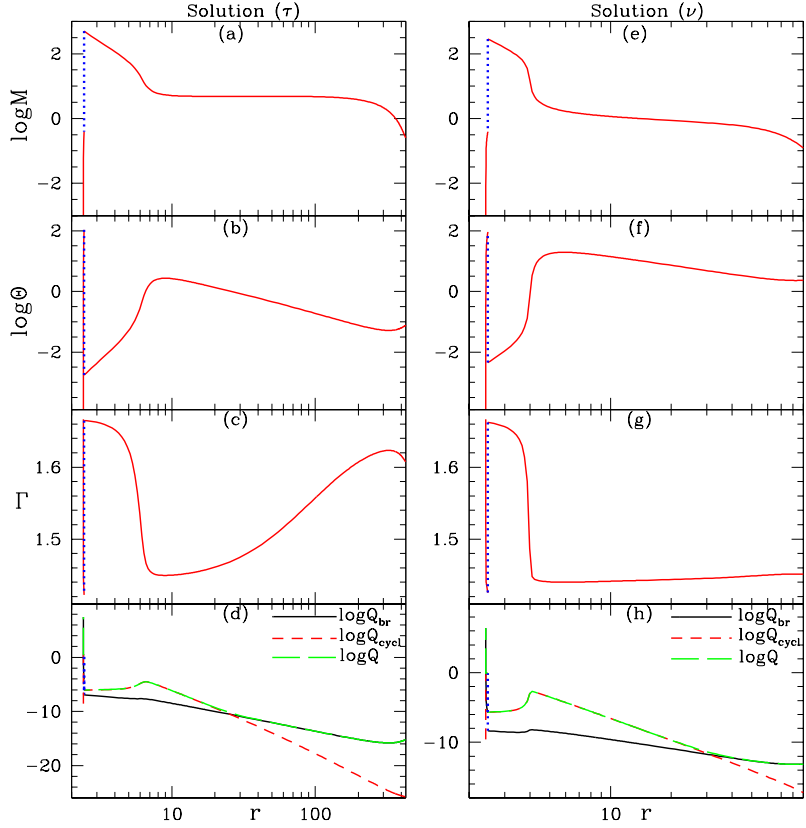, height=11truecm,width=11truecm,angle=0}
\vspace{-0.2cm}
\caption[] {\small Variation of log$M$ (a, e); log$\Theta$ (b, f); $\Gamma$ (c, g);
total radiative losses in log scale $Q$ (long dashed, green), bremsstrahlung losses $\qbr$ (solid, black)
and cyclotron losses $\qcyc$ (dashed, red) in panels (d, h) as a function of $r$. Comparison of solutions
corresponding to
coordinate points $\tau$ (a, b, c, d) and $\nu$ (e, f, g, h) of parameter space in Fig. (\ref{lab:fig3}a).
Here, $\mdd_{\tau} = 3.51\times10^{15}\mbox{g}\mbox{ s}^{-1}$, $\mdd_{\nu} = 0.35\times10^{15}\mbox{g}\mbox{ s}^{-1}$ 
and $\xi=1$.}
\label{lab:fig6}
\end{figure}

Solutions with higher and lower spins are also compared for low energy. These solutions are outside the 
MCP region. In the left panels, we plot log$M$ (Fig. \ref{lab:fig6}a), log$\Theta$ (Fig. \ref{lab:fig6}b), $\Gamma$ (Fig. \ref{lab:fig6}c) and the cooling rates log$Q$, log$\qbr$ and log$\qcyc$ (Fig. \ref{lab:fig6}d) for the solution corresponding to point $\tau$ in $\br$---$P$ space (Fig. \ref{lab:fig3}a).
In the right panels, we plot
the same set of variables (Fig. \ref{lab:fig6}e-h), but now for the parameters of the coordinate point $\nu$ in the parameter space of Fig. (\ref{lab:fig3}a).
The solution with lower spin is colder ($\tau$) than the flow with higher spin parameter ($\nu$).

The accretion solutions presented in this paper, have some interesting features. Within
a distance of $100\rg$ from the star surface, accretion streamlines
are almost radial (\ie $r$cos$\theta$=$r$sin$\theta$). However, because of the bipolar magnetic
field controls the accretion cross-section, therefore close to the stellar surface,
the cross-section is smaller than $\sim r^2$. This makes $\rho$ to be larger than a purely
radial accretion \citep[\ie Bondi accretion,][]{b52}, and consequently the cooling rates are higher
than typical Bondi type accretion. As a consequence of enhanced cooling, the temperature dips
in the region within about a $10\rg$ and the location of the terminating shock.
This dip in $\Theta$ is seen in all the temperature distributions presented above.  
In comparison, the temperature distribution of solutions without cooling do not
exhibit such dips (Fig. \ref{lab:figApp}c).
  
\begin{figure}
\hspace{3.0cm}
\psfig{figure= 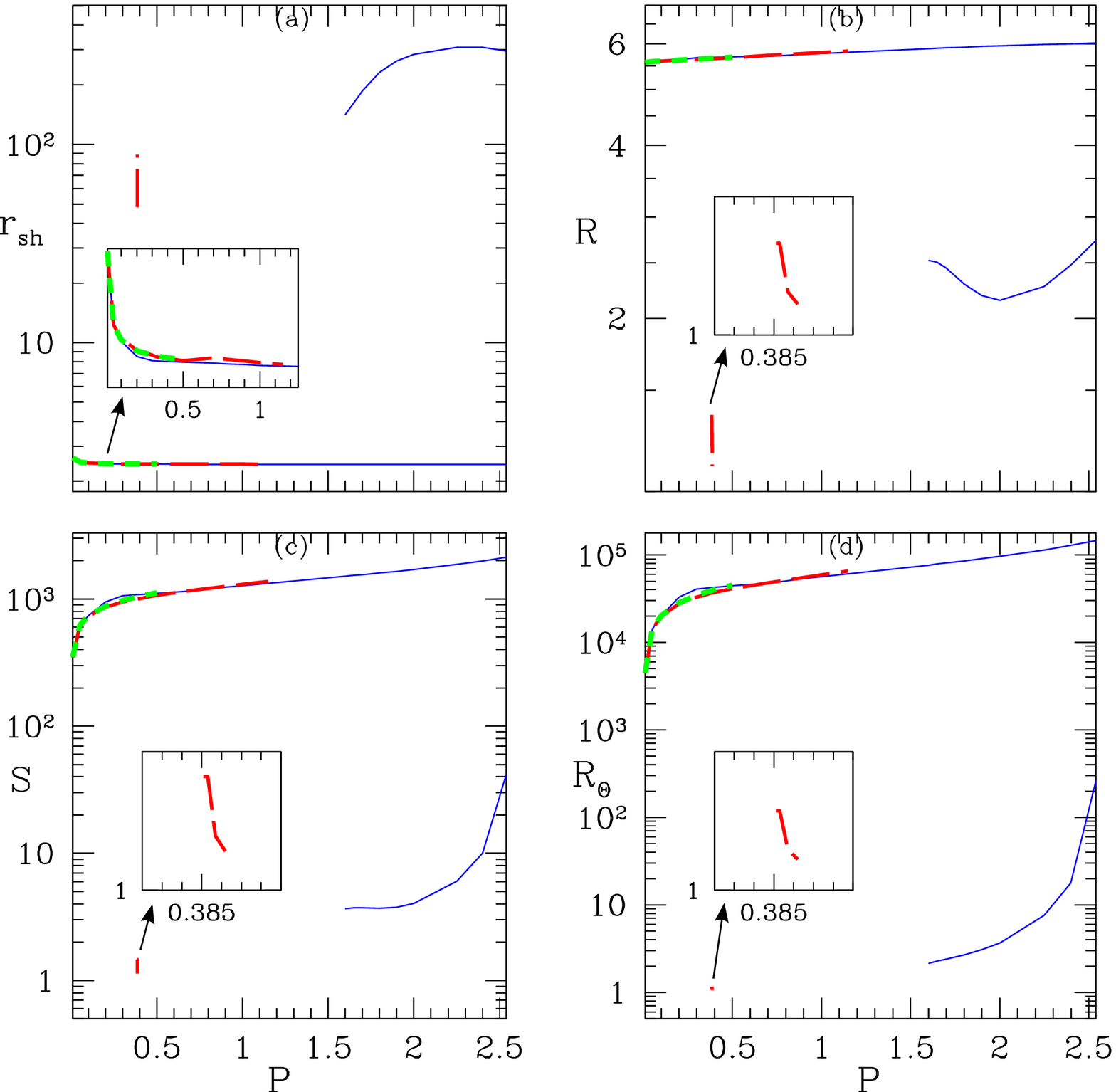,height=11truecm,width=11truecm,angle=0}
\vspace{-0.2cm}
\caption[]{\small (a) The shock location $\rsh$, (b) compression ratio $R$, (c) shock strength
$S$ and (d) temperature ratio ${\rm R}_{\Theta}$ are plotted as a function of rotation period $P$ for
$\br=0.99891$ (solid, blue), $\br=0.9983$ (long dashed, red) and  
$\br=0.997$ (dashed, green). Here, $\mdd_{0.01s} = 0.35\times10^{14}\mbox{g}\mbox{ s}^{-1}-\mdd_{2.5s} = 9\times10^{15}\mbox{g}\mbox{ s}^{-1}$
and $\xi=1$.}
\label{lab:fig7}
\end{figure}

It is clear that for a given ${\dot M}$, the properties of the accretion is determined by $\br$ and $P$.
Therefore the shock properties should also have some dependence on these two flow parameters.
In Fig. (\ref{lab:fig7}a), $\rsh$ is plotted as a function of $P$. The inner terminating shock 
is almost horizontal over-plotted curves, close to the star surface at around $2-3\rg$. The inner
shock decreases very weakly with the increase of $P$ and has almost no dependence on $\br$. The second shock is represented by the curves at few $\times 10$---$100\rg$. In Fig. (\ref{lab:fig7}b, c, d), the corresponding compression ratio $R=\rho_+/\rho_-$, shock strength $S=M_-/M_+$ and the temperature
ratio ${\rm R}_\Theta=\Theta_+/\Theta_-$ respectively, are plotted as a function of $P$. Each curve is
for $\br=0.99891$ (solid, blue), $\br=0.9983$ (long dashed, red) and  
$\br=0.997$ (dashed, green). The inner shock locations for various $\br$ are over plotted on each other, and therefore,
are zoomed in the inset of panel Fig. (\ref{lab:fig7}a). Since the parameter space for transonic solutions
are limited by the GFHD curve (Fig. \ref{lab:fig3}a), the inner shock is limited for $\br=0.997$ (dashed, green).
No second shock is found for this energy parameter for any value of $P$. For a little higher $\br=0.9983$ for a (long dashed, red), the inner shock is obtained for $P \lsim 1.1$ (see the inset). Second shock is obtained  
for a very short range of $P\sim 0.3851-0.3862$s and the range of the second shock is $48.3\leq \rsh \leq 88.5$. For $\br=0.99891$ the inner shock exist for a range of $P \lsim 2.54$ (solid, blue). The second shock is in the limited range of 
$1.6 \lsim P \lsim 2.54$, and the second shock is also located far away from the star surface
$140 \leq \rsh \leq 295$. The compression ratio $R$ (Fig. \ref{lab:fig7}b) of the inner shock is 
very high $R\sim 6$ and is almost independent of $\br$ but is very weakly dependent on $P$. The $R$ of the outer shock is $\br$ dependent.
$R\lsim 1.4$ (long dashed, red) for $\br=0.9983$ and $2 \lsim R \lsim 3$ (solid, blue) for $\br=0.99891$.
In the inset $R$ for $\br=0.9983$ is zoomed. The shock strength $S$ (Fig. \ref{lab:fig7}c) and
temperature ratio $R_\Theta$ (Fig. \ref{lab:fig7}d) plots also show
that the inner shock to be very strong and depend marginally on $\br$, but the outer shocks do depend
on $\br$ and are weak or moderate in strength. The insets in both the panels zoom the outer shocks for $\br=0.9983$. We have also calculated post-shock luminosity and found that the order of
magnitude varies with the rotation period. Therefore, the post shock luminosities at inner shock for different spin period are, ${\cal L}_{0.01s}\sim 10^{30}\mbox{erg}\mbox{ s}^{-1}$ to ${\cal L}_{2.5s}\sim 10^{35}\mbox{erg}\mbox{ s}^{-1}$ and at the outer shock is ${\cal L}\sim 10^{24-26}\mbox{erg}\mbox{ s}^{-1}$.

\begin{figure}
\hspace{3.0cm}
\psfig{figure= 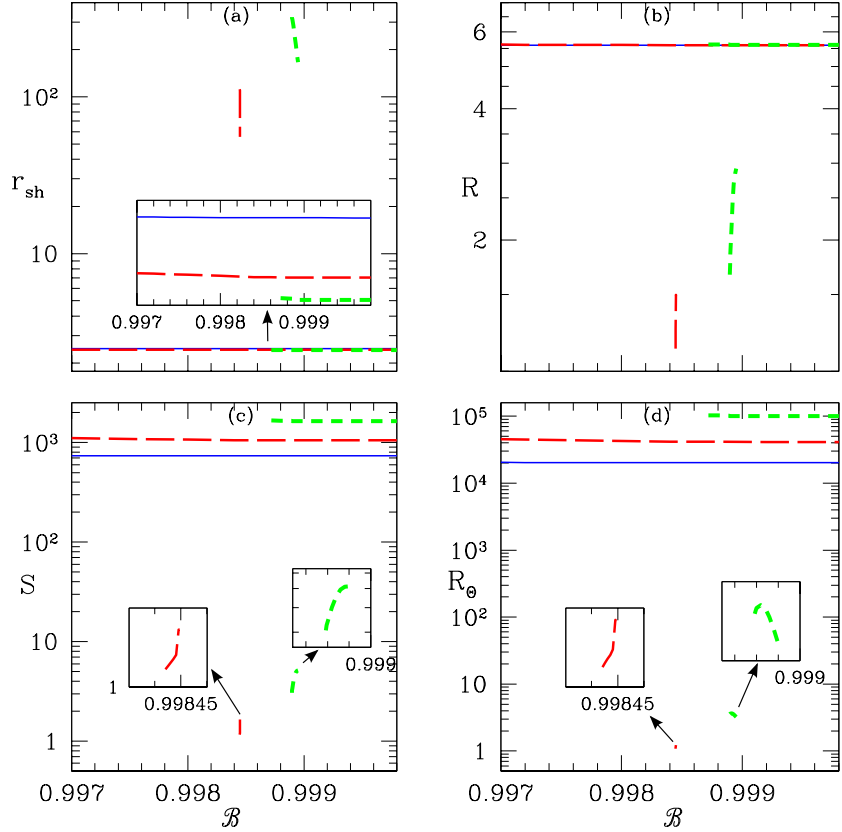,height=11truecm,width=11truecm,angle=0}
\vspace{-0.2cm}
\caption[]{\small (a) The shock location $\rsh$, (b) compression ratio $R$, (c) shock strength
$S$ and (d) temperature ratio $(\rm R_{\Theta})$ across the shock are plotted as a function of $\br$ for three rotation periods $P = 0.1$s (solid, blue), $P=0.5$s (long-dashed, red) and $P = 2.0$s (dashed, green). 
The inset in panel (a) zooms the inner shock locations. There are two insets in panels (c \& d), which zooms the outer shock quantities.  Here, $\mdd_{0.1s} = 0.35\times10^{15}\mbox{g}\mbox{ s}^{-1}$, $\mdd_{0.5s} = 1.8\times10^{15}\mbox{g}\mbox{ s}^{-1}$, $\mdd_{2.0s} = 7\times10^{15}\mbox{g}\mbox{ s}^{-1}$ 
and $\xi=1$.}
\label{lab:fig8}
\end{figure}

In Figs. (\ref{lab:fig8}a-d), all the shock variables are plotted for  $P = 0.1$s (solid, blue),
$P=0.5$s (long-dashed, red) and $P=2.0$s (dashed, green). In Fig. (\ref{lab:fig8}a), $\rsh$
is plotted as a function of $\br$. For $P=0.1$s (solid, blue) the shock is formed only close to the
star surface and for all values of $\br$. For $P=0.5$s (long-dashed, red), the inner terminating shock
forms for all values of $\br$, but at $0.998447\lsim \br \lsim 0.9984497$, outer shock forms
in a limited range $55.6 \lsim \rsh \lsim 112$. For $P=2$s (dashed, green)
inner shock forms for $\br \gsim 0.99872$. Outer shock also forms in the range $0.998896 \lsim \br \lsim
0.998949$ and the shock ranges from $166.6 \lsim \rsh \lsim 323$.
Although all the inner shock almost overlaps (lower, almost horizontal curves), the outer shock locations are perceptible.
In (\ref{lab:fig8}b), $R$ is plotted as a function of $\br$. $R \lsim 6$ for the inner shock (upper curve)
and all the curves for $P$ overlap. The compression ratio of the outer shock (lower slanted curves)
ranges from being weak to moderate. The shock strength $S$ for the inner shock depends significantly
on $P$ and are quite high. While the $S$ parameter for outer shock is comparatively much weaker
and are zoomed in the two inset panels. The $R_\Theta$ parameter for the inner shock is higher for higher
$P$ and completely dominates the outer shock (zoomed in the inset panels in Fig. \ref{lab:fig8}d).
The inner shocks are so overwhelmingly strong that the signatures of the outer shock may not be significant,
although might contribute dynamically if the outer shock are made unstable by some process. In this case, post-shock luminosities at inner shock do not change significantly with energy, and are of the order ${\cal L}_{0.1s}\sim 10^{32}\mbox{erg}\mbox{ s}^{-1}$, ${\cal L}_{0.5s}\sim 8\times 10^{33}\mbox{erg}\mbox{ s}^{-1}$, ${\cal L}_{2.0s}\sim 10^{35}\mbox{erg}\mbox{ s}^{-1}$, while at the outer shock the luminosities are ${\cal L}_{0.5s}\sim 5\times 10^{25}\mbox{erg}\mbox{ s}^{-1}$, ${\cal L}_{2.0s}\sim 10^{27}\mbox{erg}\mbox{ s}^{-1}$. 

\subsubsection{Effect of $\xi$}
\begin{figure}
\hspace{3.0cm}
\psfig{figure= 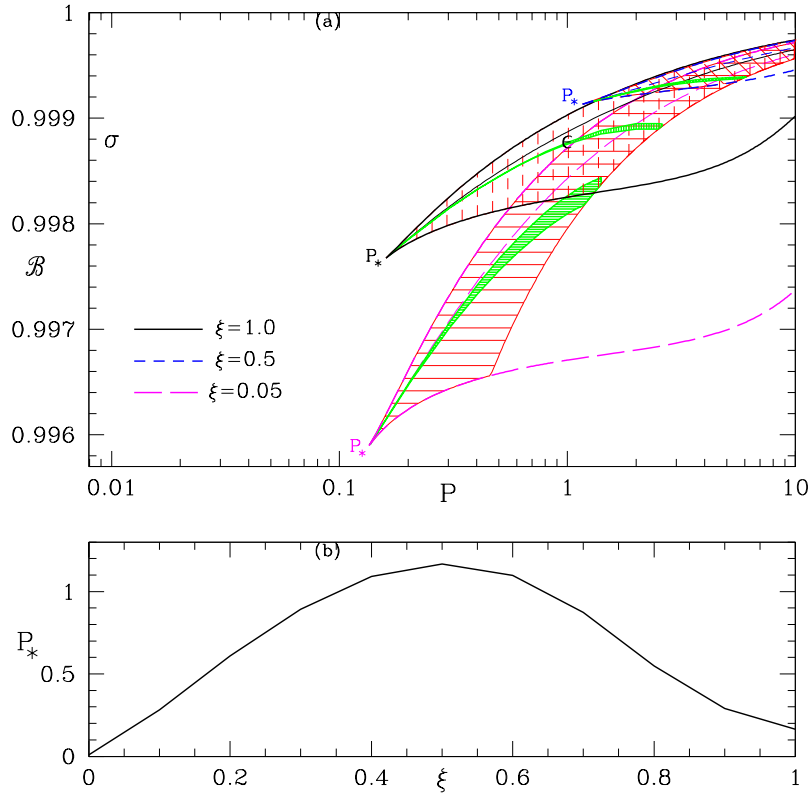,height=11truecm,width=11truecm,angle=0}
\vspace{-0.2cm}
\caption[] {\small (a) $\br$---$P$ parameter space, in which MCP region is demarcated 
for $\xi=0.05$ (long-dashed, magenta), $\xi=0.5$ (dashed, blue) and $\xi=1.0$ (solid, black).
$P_*$ is the minimum $P$ beyond which MCP is possible. Two coordinate points are marked
as `$\sigma$' and $\epsilon$', the values of $\br$, $P$ corresponding to these points
are used to obtain accretion solutions in Figs. \ref{lab:fig10} and \ref{lab:fig11}, respectively. 
(b) $P_*$ plotted as a function of $\xi$. Here, $\mdd_{0.01s} = 0.35\times10^{14}\mbox{g}\mbox{ s}^{-1}-\mdd_{10.0s} = 3.5\times10^{16}\mbox{g}\mbox{ s}^{-1}$ 
.}
\label{lab:fig9}
\end{figure}

\begin{figure}
\hspace{3.0cm}
\psfig{figure= 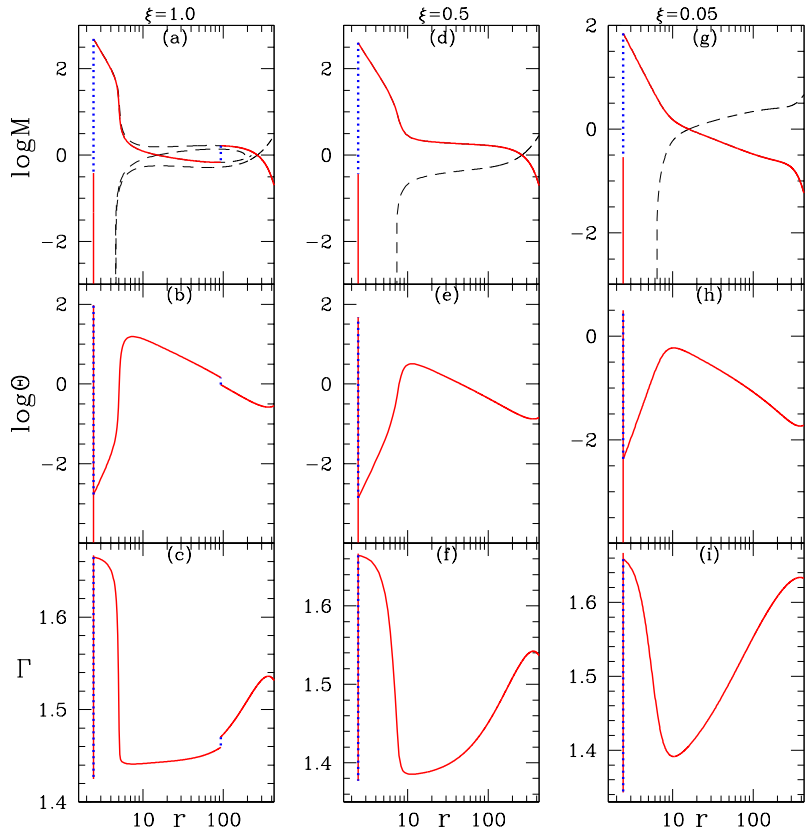,height=11truecm,width=11truecm,angle=0}
\vspace{-0.2cm}
\caption[] {\small Variation of log$M$ (a, d, g), log$\Theta$ (b, e, h) and
$\Gamma$ (c, f, i) as a function of $r$. Each column of panels
represent flow characterized by $\xi=1.0$ (a-c), $\xi=0.5$ (d-f) and $\xi=0.05$ (g-i).
The physical accretion solutions are solid curves with the shock jumps depicted as dotted (blue) vertical lines.
The crossing of the dashed and the solid curves indicate the position of the sonic points.
Here $\mdd = 3.51\times10^{15}\mbox{g}\mbox{ s}^{-1}$. 
The solutions correspond to point $\epsilon$ or $\br=0.99877$
and $P=1$s in the $\br$---$P$ parameter space of Fig. (\ref{lab:fig9}a).}
\label{lab:fig10}
\end{figure}

\begin{figure}
\hspace{3.0cm}
\psfig{figure= 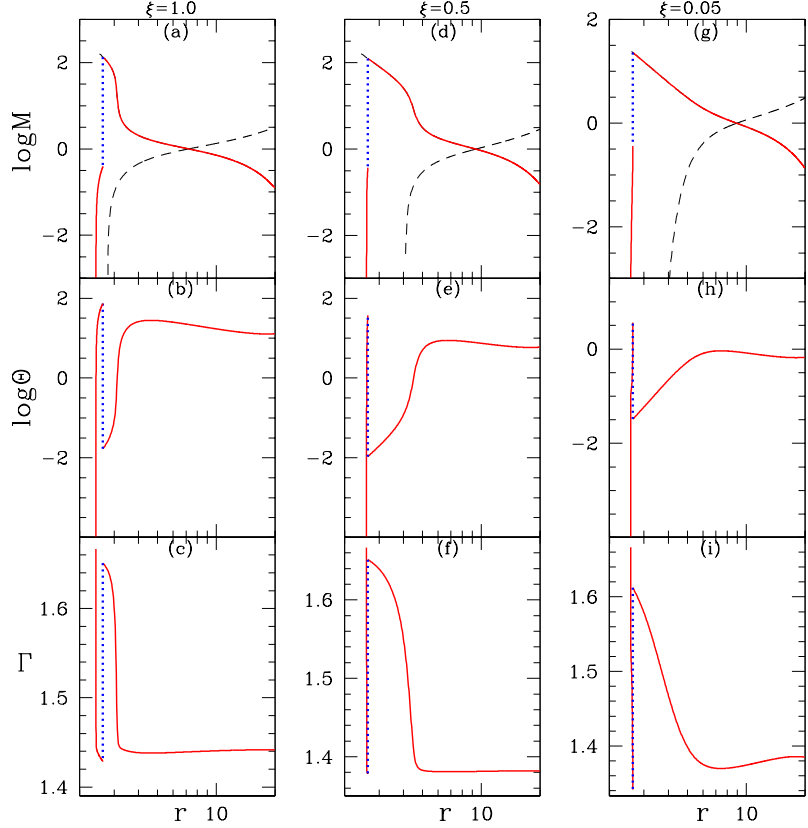,height=11truecm,width=11truecm,angle=0}
\vspace{-0.2cm}
\caption[] {\small Variation of log$M$ (a, d, g), log$\Theta$ (b, e, h) and
$\Gamma$ (c, f, i) as a function of $r$. Each column of panels
represent flow characterized by $\xi=1.0$ (a-c), $\xi=0.5$ (d-f) and $\xi=0.05$ (g-i).
The physical accretion solutions are solid curves with the shock jumps depicted as dotted (blue) vertical lines.
Here $\mdd = 0.35\times10^{14}\mbox{g}\mbox{ s}^{-1}$.
The solutions correspond to point $\sigma$ or $\br=0.99877$
and $P=0.01$s in the $\br$---$P$ parameter space of Fig. (\ref{lab:fig9}a).}
\label{lab:fig11}
\end{figure}

In Fig. \ref{lab:fig9}a, we plot the MCP region in the $\br$---$P$ parameter space for
accretion rate $\mdd_{0.01s} = 0.35\times10^{14}\mbox{g}\mbox{ s}^{-1}-\mdd_{10.0s} = 3.5\times10^{16}\mbox{g}\mbox{ s}^{-1}$, i.e., by keeping same $\rod=10^{-10}$g/cm$^3$ but for different $\xi$.
In our analysis $\xi$ is ratio of the proton to electron number density, and therefore is the composition
parameter. Each bounded region which represents MCP are for $\xi=1.0$ (solid, black), $\xi=0.5$ (dashed, blue)
and $\xi=0.05$ (long dashed, magenta), respectively.
There is a certain value of rotation period (say $P_*$) below which MCP is not possible. Small rotation period (P $<P_*$) has small $r_{co}$ or $\rd$.
Therefore, gravity is very strong in the funnel of a small $P$ accretion system. If the gravity is too
strong then, neither MCP, nor a second shock forms. However, the MCP region (area under the bounded curves) depends
significantly on $\xi$. The area under the curve shrinks for $0.5 < \xi \leq 1$ and then starts to increase.
In fact for lepton dominated flow ($\xi \sim 0.05$) the MCP region is quite significant.
This is quite different from purely hydrodynamic case. The strong magnetic field criteria increases
angular momentum at larger $r$. Therefore, in spite of the fact that the thermal energy of small $\xi$ flow
is mostly non-relativistic, but still the angular momentum is large enough at large $r$ to
modify gravity and produce multiple sonic points. 
In Fig. \ref{lab:fig8}b, we have plotted $P_*$ versus composition parameter $\xi$.
In this figure, we can see that $P_*$
starts with minimum value at $\xi=0.0$ then becomes maximum at $\xi\sim0.5$. However if $\xi$
further increases, $P_*$ starts decreasing and reaches its value at $\xi=1.0$.
Two coordinate points named as $\sigma$ ($\br=0.99877;~P=0.01$s) and $\epsilon$ ($\br=0.99877;~P=1$s)
are marked in the $\br$---$P$, chosen to 
consider high and moderate spin central stars. It may be noted that
the $\epsilon$ point is same as the coordinate point identically named in Fig. (\ref{lab:fig3}a).
Accretion solutions corresponding to these
points are compared for different $\xi$ in Figs. (\ref{lab:fig10} \& \ref{lab:fig11}).

In Figs. (\ref{lab:fig10}a-i), we have compared flow variables corresponding to $\epsilon$ in Fig. (\ref{lab:fig9}a), \ie for $\br=0.99877$ and $P=1.0$s. Each column represents solutions of flows for the same values of $\br$ and $P$, but of different composition $\xi=1.0$ (Figs. \ref{lab:fig10}a-c), $\xi=0.5$ (Figs. \ref{lab:fig10}d-f)
and $\xi=0.05$ (Figs. \ref{lab:fig10}g-i). In order to compare the solutions of different $\xi$,
in each row we have plotted log$M$ (Figs. \ref{lab:fig10}a, d, g),
log$\Theta$ (Figs. \ref{lab:fig10}b, e, h) and $\Gamma$ (Figs. \ref{lab:fig10}c, f, i).  
The same flow parameters $\br$, $P$ produces two shocks for electron-proton ($\xi=1.0$) flow, 
however, produces only the terminating shock for flows with $\xi=0.5$ and $\xi=0.05$. It is clear from Fig. (\ref{lab:fig9}a)
that $\epsilon$ is in the zone which produces multiple shocks for $\xi=1.0$, but not for $\xi=0.5$ and $\xi=0.05$.
However, for $\xi=0.5$, the point $\epsilon$ is below the MCP region, but is above the MCP region for $\xi=0.05$.
Therefore, although there is only one sonic point for both $\xi=0.5$ and $\xi=0.05$ but the sonic
points for $\xi=0.05$ is closer to the star surface than that for $\xi=0.5$.
The temperature distribution confirms conclusions from our earlier hydrodynamic
studies of multispecies flow \citep{cr09,kscc13,kc14,ck16,kc17}, \ie and electron-proton 
flow is hotter than flows dominated by leptons. 
However, although the temperature of the flow with $\xi=1.0$ is more than an order of magnitude higher than
the flow with $\xi=0.05$, but the adiabatic index distribution shows that thermally, $\xi=0.05$ flow is more relativistic than the electron-proton flow at around $r \sim 10\rg$.

In Figs. (\ref{lab:fig11}a-i), we compare the flow variables for the same $\br$ as the previous figure
but for higher spin or, $P=0.01$s and is marked in the $\br$---$P$ parameter space as the coordinate
point $\sigma$ (Fig.
\ref{lab:fig9}a). Similar to the previous figure, we plot log$M$ (Figs. \ref{lab:fig11}a, d, g),
log$\Theta$ (Figs. \ref{lab:fig11}b, e, h) and
$\Gamma$ (Figs. \ref{lab:fig11}c, f, i) as a function of $r$. Panels in each column presents distribution
of various flow variables for the same $\br $ and $P$ but for different flow composition $\xi=1.0$
(Figs. \ref{lab:fig11}a-c), $\xi=0.5$ (Figs. \ref{lab:fig11}d-f) and $\xi=0.05$ (Figs. \ref{lab:fig11}g-i).
For the parameters of $\sigma$, there are no MCP for any $\xi$ and consequently only forms the terminating shock
close to the star surface. However, the solutions differ from each other depending on $\xi$.
Apart from the difference in location of the sonic points (crossing between solid and dashed curves),
the size of the post-shock region for $\xi=1.0$ is larger than lepton dominated flows. The temperature distribution for the electron-proton flow is higher, but because of the enhanced inertia of larger fraction
of protons, $\Gamma$ shows that lepton dominated flow are thermally more relativistic than the electron-proton
flow. 

\begin{figure}
\hspace{3.0cm}
\psfig{figure= 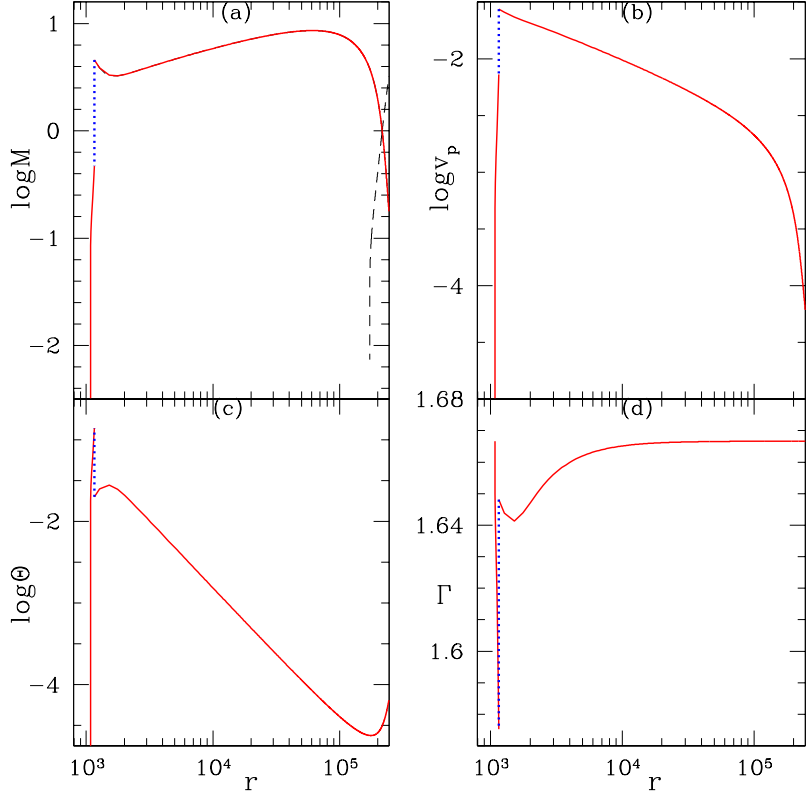,height=11truecm,width=11truecm,angle=0}
\vspace{-0.2cm}
\caption[] {\small Variation of log$M$ (a), log$\vp$ (b), log$\Theta$ (c) and
$\Gamma$ (d) as a function of $r$. 
The physical accretion solutions (solid, red), shock jumps (dotted, blue vertical lines)
and wind type solutions (dashed, black) are shown. Sonic point is at the crossing of accretion
and wind type solutions. Central object is WD with $M_{\circ}=1.2M_{\bigodot}$, $R_{\circ}=3.8\times 10^{8}$cm,
$\bpo=3\times10^{7}\mbox{G}$ and $P=12150$s. The solutions corresponds to 
$\br=0.9999968$ and $\mdd = 5.54\times10^{15}\mbox{g}\mbox{ s}^{-1}$
.}
\label{lab:fig12}
\end{figure}

\subsubsection{White Dwarf type compact object}

Among the three accepted versions of compact objects like, black hole, NS or a WD,
all have very strong gravity, although
black holes
have a very unique property of having no hard boundary and are only shielded from the outside universe
by an one way space-time screen called the event horizon. Since we are only concentrating
on magnetized accretion flow on to compact objects with hard surface, therefore black holes are beyond the scope
of this paper.
The related defining property that separates gravitation interaction that these objects impose on the surrounding
matter is the compactness parameter or $M_\circ/R_\circ$. In geometric units, the compactness
parameter for black holes $M_\circ/R_\circ = 1$; for NS it is $0.5 \lsim M_\circ/R_\circ \lsim 0.66$ and
for WDs $M_\circ/R_\circ \sim \mbox{ few}\times 10^{-4}$. Larger the compactness parameter, \ie larger the
mass packed in a finite volume, stronger is the gravity of the object, with black holes having the strongest gravity.
Up to now, we have used compactness ratio in tune
with the description of NS and all our solutions above can be thought to represent accreting NS cases.
In the following we change the central star properties to mimic accretion processes
of a WD and still using the same methodology of solution.

The mass and radius of central star used to mimic a WD is $M_{\circ}=1.2M_{\bigodot}$, $R_{\circ}=3.8\times 10^8$cm,
the surface magnetic field as $\bpo=3\times10^{7}\mbox{G}$ and the spin period $P =12150$s.
It is quite clear that a WD has quite low compactness ratio, infact in units of $\rg$, the radius of the WD
is $R_\circ=1.072\times 10^3 \rg$. Given the central star's parameters as assumed above, we need to supply 
two parameters to $\br=0.9999968$ and $\mdd = 5.54\times10^{15}\mbox{g}\mbox{ s}^{-1}$ 
we can obtain the accretion solution.
In Fig. (\ref{lab:fig12}a) we plot log$M$ as a function of $r$. The solid (red) curve represent accretion solution, the dotted (blue) vertical line represent the shock transition. The dashed line is the wind type solution (obtained with wind type boundary conditions) and its crossing with the accretion branch determines
the location of the sonic point. The accretion column terminates on the star surface after suffering the terminating shock. Other flow variables of the accretion column are log$\vp$ (Fig. \ref{lab:fig12}b), log$\Theta$
(Fig. \ref{lab:fig12}c) and $\Gamma$ (Fig. \ref{lab:fig12}d). Since the WD accretion do not achieve relativistic
temperature so we only considered electron-proton flow.
The shock location obtained from our calculations is $\rsh=1.1577 \times 10^3 \rg$, the post shock temperature is $\sim 8.313\times10^{8}$K and the post-shock density is $\sim 4.672 \times 10^{-9}\mbox{g}/\mbox{cm}^{3}$.
Therefore the shock height comes out to be $0.08R_\circ$ for the WD.
Observational studies are in general agreement with the numbers generated by us \citep{rsbb05}.
A more exact agreement can be obtained if all the dissipative processes, accretion rate
and the Bernoulli integral
are chosen properly. 
                   
\section{Discussion and Concluding Remarks}
\label{sec:conclude}
In this paper, we have studied the magnetized accretion solutions on to a compact object with hard surface.
Since the gravity of a compact object is stronger than Newtonian potential, we used PW pseudo-Newtonian
potential. Moreover, as the accretion flow traverses large distance, in the course of which the
temperature varies 2-3 orders of magnitude. So, we chose a variable $\Gamma$ or CR EoS to describe the fluid. The advantage of PW potential
is that at large distance it is essentially Newtonian, and the advantage of CR EoS is that at temperatures $< 10^7$K it behaves like fixed adiabatic index EoS with $\Gamma \sim 5/3$ \citep{cr09}. Therefore the mathematical tools we have employed will enable us to go from weak gravity to a stronger one, as well as from low to very high temperatures.
In Appendix \ref{appA}, we compare the sonic point properties of accretion flows under the influence of
Newtonian gravity and PW gravity. We ignored cooling in order to make the comparison with \citet{kol02}
more relevant. Since Newtonian gravity do not allow the formation of inner sonic point, therefore in presence of
rotation all accretion solutions possess two sonic points, the outer X-type and the inner O type (without dissipation). As a result all accretion solutions would fold back in the form of an $\alpha$ \citep[see discussions in][]{ck16,kc17}. Only some of the $\br$ and $P$ combinations would allow the turning 
radius to be less than the star radius (\eg Fig. 4a in \citet{kol02}). Only $\br<0.9981$ corresponds to a `global' solution \ie connect $\rd$ and $R_\circ$. However, in PW gravity \ie a stronger gravity, global solutions are available for all available
$\br > \bcl$. 

The velocity of the accreting matter is supersonic close to the star surface, but for a stable accretion solution
it has to either stop or corotate with the star on its surface. Since the post-shock density and temperature
is high, as well as, the presence of magnetic field, cooling process dominate and helps to minimize
the infall velocity of the accretion flow on the star surface. Therefore, in this paper all the accretion solution
ends on the star surface with a terminating shock very close to the star surface, unlike \citet{kol02,ka08}. The terminating shocks are
very strong (Figs. \ref{lab:fig7}, \ref{lab:fig8}) and therefore likely to contribute
significantly in the total electro-magnetic output of the system (Figs. \ref{lab:fig4}d,h; \ref{lab:fig5}d,f; \ref{lab:fig6}d,h).

We have calculated the total luminosity for different solutions and found that order of
magnitude is $10^{34-36} \mbox{erg} \mbox{ s}^{-1}$. Also, the electron scattering optical depth for different
solutions remains at $\lsim 0.1$ i. e., one may consider the funnel flow onto a magnetized star to be optically thin. Nonetheless, cyclotron cooling is quite complicated, and therefore is mimicked by considering a cooling function \citep{sax98,cb15}.

The effect of cooling is observed in the temperature distribution of the flow, where
enhanced cooling reduces the temperature of the flow before the terminating shock, as well
as, in the post-shock flow just above the star surface. The dip in temperature before the terminating shock
is not seen for accretion flows in which cooling processes are ignored (Fig. \ref{lab:figApp}c).

Stronger gravity of PW potential causes the formation of MCP (multiple critical
point) region in the $\br$---$P$ parameter space. This produces various accretion solutions, some of which
admits a second shock. While the terminating strong shocks, which accompanies all global accretion solutions,
are very strong $R \sim 6$ and close to the star surface $\rsh \rightarrow 2.5-3 \rg$, but second shocks
are weaker $ < 3$ and located much further out $\rsh \gsim$few$\times 10 \rg$. The electromagnetic
signature of this second shock is likely to be washed out in the steady state scenario,
but shock oscillation induced by various dissipation processes might give rise to many interesting phenomena.
The most interesting fact is that the two post shock region is
separated by a supersonic region, therefore the second shock is acoustically separated from the terminating
shock closer to the star surface, although the oscillations in the second shock in principle can make the terminating
shock time dependent.
How this will pan out in terms of observation, needs to be determined through numerical simulations
and is currently beyond the scope of this paper.

Although there is a general agreement on various features with hydrodynamic black hole accretion
solutions, but there are some significant differences as well. The cross-sectional area of the
accretion is smaller than a typical $\sim r^2$ cross section expected in the inner
region of a black hole
accretion disc. Therefore, the density of matter near a NS or WD surface is much larger than the one
near black hole horizon.
As a result bremsstrahlung, cyclotron and other cooling processes are much more important near the star surface
than a black hole.

Another important difference is the sonic point properties. In the hydrodynamic case, $\rc$ can be as large as possible and for $\rc \rightarrow$large, $\brc~\rightarrow 1$. However, in the present case maximum $\rc$
possible is $r_{\rm cl}$ and the corresponding energy is $\bcl$. So, in case of purely hydrodynamic
flow X-type sonic points for global solutions, are obtained if $\br > 1$, but here X-type sonic points
related to a global solution are obtained if $\br > \bcl$.
Since CR EoS also contains the information of composition, therefore we also studied how the proton
content affects the solution. In the purely hydrodynamic case, the proton poor flow is thermally non-relativistic
and the MCP region is small and vanishes for $\xi=0$. But in the present case, the MCP region is large for proton
poor flow. This is because, the strong magnetic field decreases angular momentum close to the
star, but increases at larger distance. So whatever may be the thermal state of the flow,
centrifugal interaction alone is important enough at large distance to modify gravity and produce multiple sonic points. 
So even for $\xi \sim 0$ flow, multiple sonic points and second shocks
are possible in presence of magnetic field. 

In conclusion, we obtained solutions of accretion flow onto a magnetized compact star, in presence of bremsstrahlung
and cyclotron cooling, in the strong magnetic field approximation. We also used variable $\Gamma$ EoS to describe
the fluid. All global accretion solution forms a very strong terminating shock near the star surface. This terminating shock also contributes significantly to the total radiation budget of these accretion systems. There
is also a possibility of forming multiple sonic points in a limited range of flow parameters. A weak to moderately strong shock at a distance of about hundred $\rg$ is formed for flows characterized by the parameters
from this limited range of parameters. Accretion flow
with two shocks are a class of
interesting solutions and its implication need to be investigated further. The overall luminosity
obtained, ranged between $10^{34-36}$ergs s$^{-1}$ and the efficiency is little more than $0.1$. Same methodology can be employed
for NS or WD type compact objects. A typical accretion solution onto a WD type compact object, generates
satisfactory post-shock properties. 

\section*{Acknowledgment}

The authors acknowledge the anonymous referee for helpful suggestions to improve the quality of the paper.
The authors also acknowledge Dipankar Bhattacharya and Prasanta Bera for important discussions on the topic
of accretion on to magnetized stars.

\appendix
\section{Comparison between Newtonian and PW gravity and the different EoS}
\label{appA}
Here we compare solutions obtained by assuming (I) Newtonian gravitational potential and
fixed $\Gamma$ EoS of the flow, with those obtained by using (II) PW gravitational potential and CR EoS.
It is to be remembered that the solutions of
\citet{kol02} are the reference for type I solutions. As has been mentioned in section \ref{sec:intro},
the solutions of \citet{kol02} ignore cooling, so to compare we also ignored
cooling for both I and II type solutions.
The mathematical structure and solution methodology is exactly same as that mentioned in the
main text, except that while the EoS of type II solution is given by equation (\ref{etrnl.eq})
from section \ref{subsec:eos}, but for type I solution the EoS is
\begin{equation}
{\bar e}=\frac{p}{\Gamma -1};~\mbox{ here } \Gamma \mbox{ is fixed!}
\label{eosfixed.eq}
\end{equation}
The entropy-accretion rate for type II solution is given by equation (\ref{ent_rate.eq}), but for
type I solution, it is given by
\begin{equation}
\md=c_s^{2/(\Gamma -1)}\vp \ap
\label{mdotfixed.eq}
\end{equation}
Since the equation (\ref{eosfixed.eq}) is quite different than CR EoS, and moreover do not
contain the information of rest mass energy density therefore, the values of Bernoulli parameter
as well as, the entropy accretion rate are quite different.
Besides, the unit system chosen by \citep{kol02} is also different from this paper.
The unit of velocity chosen in this paper and type II solutions is $c$, while that of type I and \citet{kol02}
is $\Omega \rd$.
To convert of $\br$ from \citet{kol02} to ours, we first obtain $\br$ in terms of physical units
and then divide that with $c^2$. Moreover, since EoS of type I solutions do not contain the 
information of rest energy density we added it to make it comparable to the $\br$ of type II solutions.
Furthermore, similar to Fig. 4a
of \citet{kol02}, for type I solutions we choose $\Gamma=5/3$ as the representative case.

\begin{figure}
\hspace{3.0cm}
\psfig{figure= 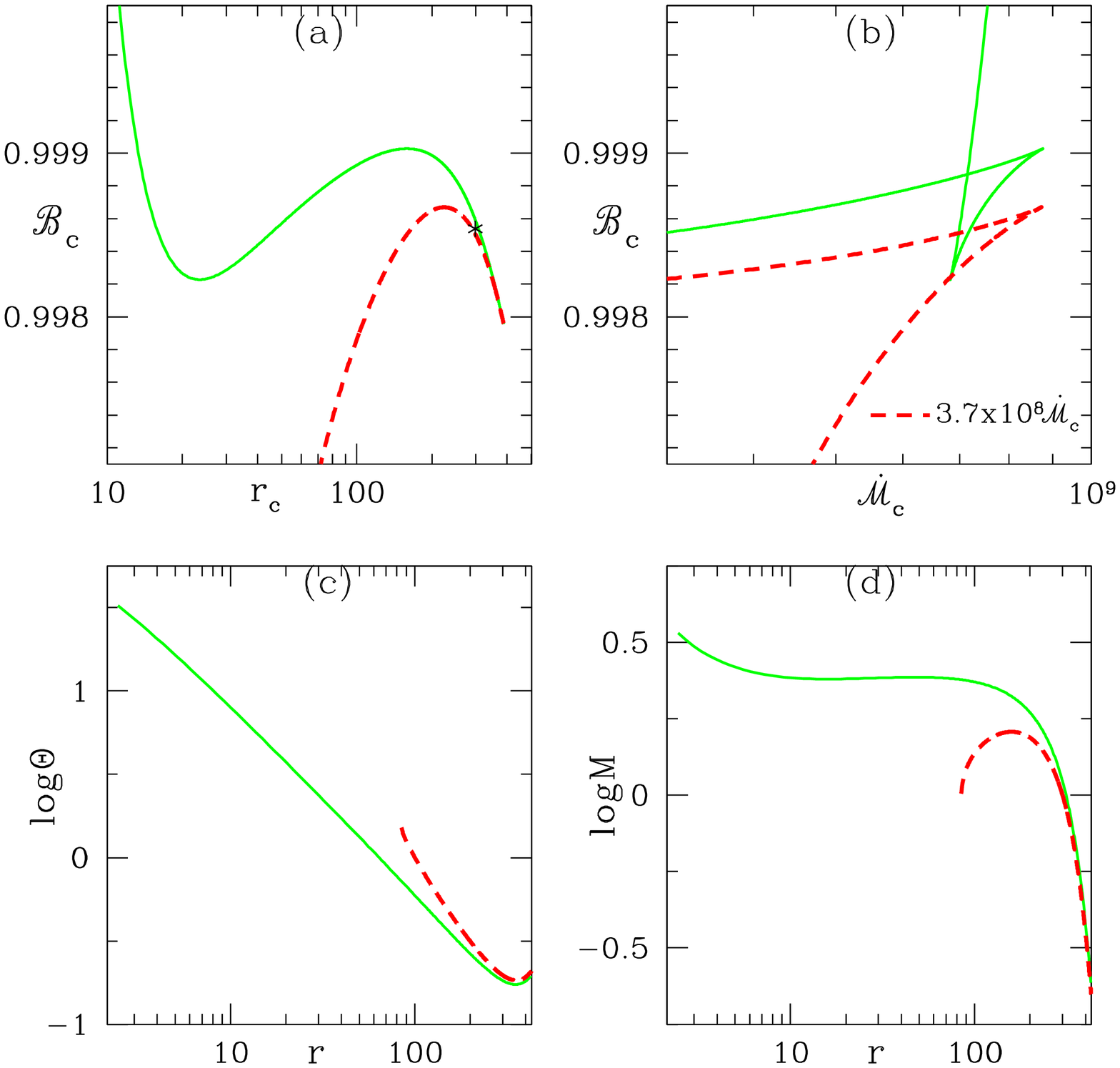,height=11truecm,width=11truecm,angle=0}
\vspace{-0.2cm}
\caption[] {\small Sonic point properties: (a) $\brc$---$\rc$ and (b) $\brc$---${\dot {\cal M}}_c$.
Solutions: (c) log$\Theta$, (d) log$M$ as a function of $r$ for flow parameters
$\br=0.9985$ and $P=1$s (star mark in panel a). Each curve compares type I solution
\ie fixed $\Gamma~(=5/3)$ EoS and Newtonian potential (dashed, red) and type II or those with CR EoS and PW
gravitation potential (solid, green).
In both the solutions cooling is ignored.}
\label{lab:figApp}
\end{figure}

In Fig. (\ref{lab:figApp}a) we plot $\brc$ as a function of $\rc$ for an NS of $P=1$s. For the type I
case (dashed, red), there is only a maximum, but for type II solution there is a maximum and a minimum.
Therefore for a given $\br=\brc$ within the maximum and the minimum value, there can be three sonic points for type II solution, but only two sonic points for the type I case.
This is the effect of PW potential over the Newtonian version. A stronger gravity ensures the formation of
the inner sonic point. The star mark corresponds to the value $\br=0.9985$.
Figure (\ref{lab:figApp}b) reconfirms the same fact that the inner sonic point does not form for type I solutions.
In order to make the entropy-accretion rate of type I solutions comparable with that of the type II,
a large factor is multiplied with the former. 
Sonic point properties are fundamentally different between type I and II cases. For $P=1$s, type I solutions
have limited range of $\br$ which connect $\rd$ and $R_\circ$. But for type II case, global solutions
connecting $\rd$ and $R_\circ$ can be obtained for all available $\br$. In Fig. (\ref{lab:figApp}c)
we compare log$\Theta$ with $r$ for type I (dashed, red) and type II (solid, green) solutions.
Type I solution terminates before reaching the star surface. In Fig. (\ref{lab:figApp}d)
we compare log$M$ with $r$. Since type I solution has two sonic points, an outer X type and a middle O type,
therefore the accretion solution does not reach the star ($M \rightarrow 0$ at $r>R_\circ$).
Type II solution is global. Since \citet{kol02} did not compute shocks, so we just present
the transonic solution. The $M$ distribution of type II solution indicates the presence of middle
and inner sonic points (regions where $dM/dr \rightarrow 0$). Since we have ignored cooling in the results
presented in this appendix, therefore, we see the temperature monotonically increases inward.

{}
\end{document}